\documentclass[12pt]{article}
\usepackage{amsmath,amsfonts,graphicx,color,bbm,tikz}
\usepackage{amssymb}
\usepackage{here}

\usepackage{bm}
\usepackage{tikz}
\usepackage{caption}
\usepackage[nosort]{cite}

\newcommand{\be}{\begin{equation}}
\newcommand{\ee}{\end{equation}}
\newcommand{\bea}{\begin{eqnarray}}
\newcommand{\eea}{\end{eqnarray}}
\newcommand{\Tr}{{\rm Tr}}
\newcommand{\tr}{{\rm tr}}
\newcommand{\rF}{{\rm F}}
\newcommand{\rM}{{\rm M}}

\newcommand{\cM}{\mathcal{M}}

\newcommand{\id}{{\bf 1}}
\def\bC {\mathbb{C}}

\def\bZ {\mathbb{Z}}
\newcommand{\vev}[1]{{\left\langle {#1} \right\rangle}}
\newcommand{\bra}[1]{{\left\langle {#1} \right|}}
\newcommand{\ket}[1]{{\left| {#1} \right\rangle}}

\def\nn{\nonumber}
\newcommand{\Slash}[1]{{\ooalign{\hfil/\hfil\crcr$#1$}}}
\newcommand{\binomi}[2]{\begin{pmatrix} #1 \\ #2 \end{pmatrix}}

\begin{document}
\thispagestyle{empty} \addtocounter{page}{-1}
\begin{flushright}
%
\end{flushright} 
\vspace*{1cm}

\begin{center}
{\large \bf Highly entangled spin chains and 2D quantum gravity}\\
\vspace*{2cm}
Fumihiko Sugino\\
\vskip0.7cm
{\it Center for Theoretical Physics of the Universe, Institute for Basic Science (IBS)} \\
\vspace*{1mm}
{\it Daejeon 34126, Republic of Korea}\\
\vspace*{0.2cm}
{\tt fusugino@gmail.com}\\
\end{center}
\vskip1.5cm
\centerline{\bf Abstract}
\vspace*{0.3cm}
{\small 
Motzkin and Fredkin spin chains exhibit the extraordinary amount of entanglement scaling as a square-root of the volume, which is beyond 
logarithmic scaling in the ordinary critical systems. Intensive study of such spin systems is urged to reveal novel features of quantum entanglement.   
As a study of the systems from a different viewpoint, we introduce large-$N$ matrix models with so-called $ABAB$ interactions, 
in which correlation functions reproduce the entanglement scaling in tree and planar Feynman diagrams. 
Including loop diagrams naturally defines an extension of the Motzkin and Fredkin spin chains. 
Contribution from the whole loop effects at large $N$ 
gives the growth of the power of $3/2$ (with logarithmic correction), further beyond the square-root scaling. 
The loop contribution provides fluctuating two-dimensional bulk geometry, 
and the enhancement of the entanglement is understood as an effect of quantum gravity. 
}
\vspace*{1.1cm}



\newpage

\section{Introduction}
\label{sec:intro}
\setcounter{equation}{0}
Entanglement is one of the most characteristic features of quantum mechanics, which provides correlations between objects that are unexplainable within classical mechanics. 
For a given system $S$ that is supposed to be divided into two subsystems $A$ and $B$, the reduced density matrix of $A$ is defined by tracing out the degrees of freedom of $B$ 
in the density matrix of the total system $\rho_S$:
$ 
\rho_A=\Tr_B\rho_S$,
where $\Tr_B$ means the trace over the Hilbert space belonging to $B$. 
Even if $\rho_S$ is a pure state, i.e., can be expressed as the form $\rho_S=\ket{\psi}\bra{\psi}$ for some state $\ket{\psi}$, 
$\rho_A$ is no longer so in general and takes a form like $\rho_A=c_1\ket{\psi_1}\bra{\psi_1} + c_2\ket{\psi_2}\bra{\psi_2}+\cdots$ ($c_i$'s are positive numbers summed to 1) 
that is called a mixed state. 
As a measure of the entanglement, the entanglement entropy (EE) is defined as the von Neumann entropy with respect to $\rho_A$: 
\be
S_A=-\Tr(\rho_A\ln\rho_A) , 
\label{vNS}
\ee
which vanishes for the pure states but not for the mixed states. 
$\rho_A$ carries information of interactions between $A$ and $B$, some of which can be read off through (\ref{vNS}).  
We can say that difference of the behavior of (\ref{vNS}) reflects difference of dynamical property of the system. 

Let us consider ground states of quantum many-body systems with local interactions. Normally, their EEs are proportional to the area of the boundaries of $A$ and $B$ 
(called as area law~\cite{eisert}).  
This can be naturally understood in gapped systems because the correlation length is finite and relevant interactions to the EE are localized along the boundaries. 
However, gapless systems are exceptional. For example, in $(1+1)$-dimensional conformal field theory, the EE violates the area law by a logarithmic factor, 
namely grows as the logarithm of the volume of the subsystem~\cite{wilczek,korepin,calabrese}. Recently, Movassagh and Shor discovered a quantum spin chain (called as Motzkin spin chain), 
whose EE grows as a square-root of the volume and greatly violates the area law in spite of local interactions~\cite{motzkin}. 
A different spin chain with smaller degrees of freedom but exhibiting the same scaling of the EE, 
called as Fredkin spin chain, was constructed by Salberger and Korepin~\cite{fredkin,dellanna_etal}.  

In this paper, we introduce large-$N$ matrix models whose correlation functions at the tree and planar level reproduce the square-root scaling of the EEs of the Motzkin and Fredkin spin chains. 
By including loop contribution, such matrix models naturally give an extension of the spin chains. By analyzing the exact solution of one of the matrix models, 
we find that analogous quantity to the EE including loop effects scales as the power of $3/2$ (with logarithmic correction) beyond the square-root.  
Whereas the tree diagrams are called as rainbow diagrams and look like skeletons~\cite{rainbow}, loop effects generate 
diagrams like fishnets that dominate around a critical point and can be regarded as a random surface. This gives intuitive understanding of the enhancement of the correlation and the entanglement 
between the subsystems. 
Since the emerging random surface picture defines quantum gravity on two-dimensional bulk, it would be intriguing to discuss the models from the holographic point of view.   

This paper is organized as follows. In sections~\ref{sec:fredkin} and \ref{sec:motzkin} the Fredkin and Motzkin spin chains and their EEs of ground states are briefly reviewed. 
In section~\ref{sec:MM}, we introduce large-$N$ matrix models and their connection to the Fredkin and Motzkin spin chains is discussed. 
In section~\ref{sec:MM_sol}, from the exact solution of one of the matrix models, we compute analog of the EE that includes effects of fluctuating bulk geometry, 
and find the enhancement of the square-root scaling to the power of $3/2$. 
Section~\ref{sec:summary} is devoted to summarize the result and discuss some future directions. 
The matrix models have so-called $ABAB$ interactions, which are not soluble in the standard manner. In appendix~\ref{app:MM_sol}, we briefly explain the exact solution 
obtained by character expansion in~\cite{KZJ}. 
Based on the solution, we compute more nontrivial one-point functions from Schwinger-Dyson (SD) equations in appendix~\ref{app:SDeqs}, 
which are used in section~\ref{sec:MM_sol}.

\section{Fredkin spin chain}
\label{sec:fredkin}
\setcounter{equation}{0}
We start with a spin chain of length $2n$, where up and down spin degrees of freedom with multiplicity (called as color) $s$ are assigned at each of the lattice sites $\{1,2,\cdots, 2n\}$. 
The up- and down-spin states with color $k\in \{1,2,\cdots,s\}$ at the site $i$ is expressed as $\ket{u^k_i}$  and $\ket{d^k_i}$, respectively.  
The Hamiltonian of the Fredkin spin chain~\cite{fredkin,dellanna_etal} is given by the sum of projection operators:
\begin{align}
H_{\rF, s} & =   \sum_{j=1}^{2n-2}\sum_{k_1,k_2,k_3=1}^s \left\{\ket{U^{k_1,k_2,k_3}_{j, j+1, j+2}}\bra{U^{k_1,k_2,k_3}_{j, j+1, j+2}}+
\ket{D^{k_1,k_2,k_3}_{j, j+1, j+2}}\bra{D^{k_1,k_2,k_3}_{j, j+1, j+2}}\right\} \nn \\
&  +\sum_{j=1}^{2n-1} \sum_{k\neq \ell}\left\{\ket{u^k_j, d^\ell_{j+1}}\bra{u^k_j, d^\ell_{j+1}}
+\frac12\left(\ket{u_j^k, d^k_{j+1}}-\ket{u^\ell_j, d^{\ell}_{j+1}}\right)\left(\bra{u_j^k, d^k_{j+1}}-\bra{u^\ell_j, d^{\ell}_{j+1}}\right)\right\} \nn \\
&  + \sum_{k=1}^s\left\{\ket{d^k_1}\bra{d^k_1} + \ket{u^k_{2n}}\bra{u^k_{2n}}\right\},
\label{HF}
\end{align}
where
\bea
\ket{U^{k_1,k_2,k_3}_{j, j+1, j+2}} & = & \frac{1}{\sqrt{2}}\left(\ket{u^{k_1}_j, u^{k_2}_{j+1}, d^{k_3}_{j+2}}
-\ket{u^{k_1}_j, d^{k_2}_{j+1}, u^{k_3}_{j+2}}\right),\nn  \\
\ket{D^{k_1,k_2,k_3}_{j, j+1, j+2}} & = & \frac{1}{\sqrt{2}}\left(\ket{u^{k_1}_j, d^{k_2}_{j+1}, d^{k_3}_{j+2}}
-\ket{d^{k_1}_j, u^{k_2}_{j+1}, d^{k_3}_{j+2}}\right). 
\eea
The Hamiltonian consists of local interactions ranging up to next-to-nearest neighbors. 

For colorless case ($s=1$), the up- and down-spin states can be represented as arrows in the two-dimensional $(x,y)$-plane pointing to $(1,1)$ (up-step) and $(1,-1)$ (down-step), respectively. 
Then, a spin configuration of the chain corresponds to a length-$2n$ path consisting of the up- and down-steps. The Hamiltonian (\ref{HF}) has a unique ground state at zero energy, which 
is superposition of spin configurations with equal weight. Each spin configuration appearing in the superposition is identified with each path of length-$2n$ Dyck walks that are random walks starting 
at the origin, ending at $(2n,0)$, and restricted to the region $y\geq 0$. 

For $s$-color case, the above identification is still valid with each spin state and its corresponding step being endowed with color degrees of freedom. 
Each spin configuration of the chain corresponds to a length-$2n$ path that consists of the up- and down-steps with colors. 
The ground state is unique, and corresponds to length-$2n$ colored Dyck walks, in which the color of each up-step should be matched with that of the down-step 
subsequently appearing at the same height. 
The other is the same as the colorless case.         

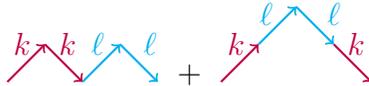
\begin{figure}[H]
\centering
\captionsetup{width=.8\linewidth}
\begin{tikzpicture}
\draw[purple, ->, thick] (0,0)--(0.5,0.5);
\draw[purple, ->, thick] (0.5,0.5)--(1,0);
\draw[cyan, ->, thick] (1,0)--(1.5,0.5);
\draw[cyan, ->, thick] (1.5,0.5)--(2,0);
\node (k) at (0.2,0.5) {\textcolor{purple}{$k$}};
\node (k) at (0.8,0.5) {\textcolor{purple}{$k$}};
\node (k') at (1.2,0.5) {\textcolor{cyan}{$\ell$}};
\node (k') at (1.9,0.5) {\textcolor{cyan}{$\ell$}};
\end{tikzpicture} 
$+ \,$
\begin{tikzpicture}
\draw[purple, ->, thick] (0,0)--(0.5,0.5);
\draw[cyan, ->, thick] (0.5,0.5)--(1,1);
\draw[cyan, ->, thick] (1,1)--(1.5,0.5);
\draw[purple, ->, thick] (1.5,0.5)--(2,0);
\node (k) at (0.2,0.5) {\textcolor{purple}{$k$}};
\node (k') at (0.6,0.9) {\textcolor{cyan}{$\ell$}};
\node (k') at (1.5,0.9) {\textcolor{cyan}{$\ell$}};
\node (k) at (1.8,0.5) {\textcolor{purple}{$k$}};
\end{tikzpicture}
\caption{Colored Dyck walks in the summand of (\ref{PF4_Dyck}). Up- and down-steps with the same color are matched.}
\label{fig:PF4}       
\end{figure}

The ground state is given by 
\be
\ket{P_{\,\rF,\,2n,\,s}}=\frac{1}{\sqrt{N_{\rF,\,2n,\,s}}}\sum_{w\in P_{\,\rF,\,2n,\,s}}\ket{w},
\ee
where $P_{\,\rF,\, 2n,\,s}$ denotes the formal sum of length-$2n$ colored Dyck walks, 
$w$ runs over monomials appearing in $P_{\,\rF,\,2n,\,s}$, and 
$N_{\rF,\,2n,\,s}$ stands for the number of the length-$2n$ colored Dyck walks: 
\be
N_{\rF,\,2n,\,s}=s^n\,N_{\rF,\,2n}=\frac{s^n}{n+1}\binomi{2n}{n}.
\label{N2n_F}
\ee
$N_{\rF,\,2n}$ denotes the number 
of colorless Dyck walks of length $2n$, which is equal to the $n$-th Catalan number. 
$N_{\rF,\,2n,\,s}$ can be obtained by setting all the $u^k$ and $d^k$ to 1 in $P_{\,\rF,\,2n,\,s}$. 
For example, $2n=4$ case reads 
\bea
P_{\,\rF,\,4,\,s} & = & \sum_{k,\ell=1}^s \left(u^kd^ku^\ell d^\ell + u^k u^\ell d^\ell d^k\right), 
\label{PF4_Dyck}\\
\ket{P_{\,\rF,\,4,\,s}} & = & \frac{1}{\sqrt{2s^2}}\sum_{k,\ell=1}^s
\left\{\ket{u^k_1, d^k_2, u^\ell_3, d^\ell_4} + \ket{u^k_1, u^\ell_2, d^\ell_3,d^k_4}\right\}. 
\label{PF4}
\eea
The two states of the summand are drawn as colored Dyck walks in Fig.~\ref{fig:PF4}. 
%


\subsection{EE of the ground state}
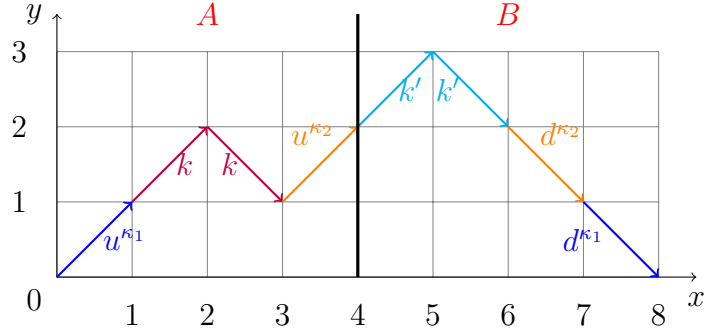
\begin{figure}[H]
\centering
\captionsetup{width=.8\linewidth}
\begin{tikzpicture}
\draw[help lines] (0,0) grid (8,3);
\node (x) at (8.5,-0.3) {$x$};
\node (y) at (-0.3,3.5) {$y$};
\node (0) at (-0.3,-0.3) {0};
\node (1) at (1,-0.5) {1};
\node (2) at (2,-0.5) {2};
\node (3) at (3,-0.5) {3};
\node (4) at (4,-0.5) {4};
\node (5) at (5,-0.5) {5};
\node (6) at (6,-0.5) {6};
\node (7) at (7,-0.5) {7};
\node (8) at (8,-0.5) {8};
\node (1) at (-0.5,1) {1};
\node (2) at (-0.5,2) {2};
\node (3) at (-0.5,3) {3};
\draw[<->] (0,3.5)--(0,0)--(8.5,0);
\draw[blue,->, thick] (0,0)--(1,1);
\draw[purple, ->, thick] (1,1)--(2,2);
\draw[purple, ->, thick] (2,2)--(3,1);
\draw[orange,->, thick] (3,1)--(4,2);
\draw[cyan, ->, thick] (4,2)--(5,3);
\draw[cyan, ->, thick] (5,3)--(6,2);
\draw[orange,->, thick] (6,2)--(7,1);
\draw[blue,->, thick] (7,1)--(8,0);
\node (k) at (1.7,1.5) {\textcolor{purple}{$k$}};
\node (k) at (2.3,1.5) {\textcolor{purple}{$k$}};
\node (k') at (4.7,2.5) {\textcolor{cyan}{$k'$}};
\node (k') at (5.2,2.5) {\textcolor{cyan}{$k'$}};
\node (u1) at (0.9,0.5) {\textcolor{blue}{$u^{\kappa_1}$}};
\node (d1) at (7.0,0.5) {\textcolor{blue}{$d^{\kappa_1}$}};
\node (u2) at (3.4,1.9) {\textcolor{orange}{$u^{\kappa_2}$}};
\node (d2) at (6.7,1.9) {\textcolor{orange}{$d^{\kappa_2}$}};
\draw[very thick] (4,3.5)--(4,0);
\node (A) at (2,3.5) {\textcolor{red}{$A$}};
\node (B) at (6,3.5) {\textcolor{red}{$B$}};
\end{tikzpicture}
\caption{A path in case of $2n=8$, $r=0$ and $h=2$.  Colors $k$ and $k'$ are matched in $A$ itself and in $B$ itself, respectively. On the other hand, 
colors of $\kappa_1$ and $\kappa_2$ are unmatched in $A$ or $B$ alone, but matched across the boundary of $A$ and $B$.}
\label{fig:PF8}       
\end{figure}

We divide the total system into two subsystems (called as $A$ and $B$), and compute the EE by tracing out spins in $B$. 
Here, let us take a block of the first $(n+r)$ spins as $A$ and the remaining $(n-r)$ spins as $B$, and consider the case $n\pm r=O(n)\to \infty$. 
Spin configurations in $A$ correspond to a part of colored Dyck paths from the origin to $(n+r, h)$ in the $(x,y)$-plane, denoted by $P^{(0\to h)}_{\rF, \,n+r,\,s}$. 
The height $h$ takes non-negative integers. Similarly, spin configurations in $B$ correspond to the paths from $(n+r, h)$ to $(2n,0)$, denoted by $P^{(h\to 0)}_{\rF,\, n-r,\,s}$. 
Note that for any colored Dyck path, the part $P^{(0\to h)}_{\rF, \,n+r,\,s}$ has $h$ unmatched up-steps that are supposed to be matched across the boundary with $h$ unmatched down-steps in  
the part $P^{(h\to 0)}_{\rF,\, n-r,\,s}$. 
Let  $\tilde{P}^{(0\to h)}_{\rF,\,n+r,\,s}(\{\kappa_m\})$ 
($\tilde{P}^{(h\to 0)}_{\rF,\,n-r,\,s}(\{\kappa_m\})$) be 
paths belonging to $A$ ($B$) with colors of the unmatched up- (down-) steps fixed to $\kappa_1, \cdots, \kappa_h$, 
where $\kappa_m$ denotes the color of unmatched up- or down-step connecting the heights 
$m-1$ and $m$. 
An example of a path in case of $2n=8$, $r=0$ and $h=2$ is depicted in Fig.~\ref{fig:PF8}.
%

Combinatorial arguments give the numbers of the paths $P^{(0\to h)}_{\rF, \,n+r,\,s}$ and $\tilde{P}^{(0\to h)}_{\rF,\,n+r,\,s}(\{\kappa_m\})$ as 
\be
N_{\rF,\, n+r,\, s}^{(0\to h)} 
= s^{\frac{n+r+h}{2}}N^{(h)}_{\rF,\,n+r} 
\quad \mbox{and} \quad
\tilde{N}_{\rF,\,n+r,\,s}^{(0\to h)} = s^{-h} N_{\rF,\, n+r,\, s}^{(0\to h)} =s^{\frac{n+r-h}{2}}N^{(h)}_{\rF,\,n+r}
\label{Ntilde_F}
\ee
with 
\be
N^{(h)}_{\rF,\,n+r}=\frac{1+(-1)^{n+r+h}}{2}\frac{h+1}{\frac{n+r+h}{2}+1}\binomi{n+r}{\frac{n+r+h}{2}}.
\label{Nhn_F}
\ee
It is easy to see that $N_{\rF,\, n-r,\, s}^{(h\to 0)} =N_{\rF,\, n-r,\, s}^{(0\to h)}$ and 
$\tilde{N}_{\rF,\,n-r,\,s}^{(h\to 0)}=\tilde{N}_{\rF,\,n-r,\,s}^{(0\to h)}$.   
The ground state is decomposed as a linear combination of tensor products of two states belonging to $A$ and $B$ 
(Schmidt decomposition):
\be
\ket{P_{\,\rF,\,2n,\,s}} = \sum_{h=0}^{n-|r|}\sum_{\kappa_1=1}^s\cdots \sum_{\kappa_h=1}^s \sqrt{p^{(h)}_{\rF,\,n+r,n-r,\,s}}\,
\ket{\tilde{P}^{(0\to h)}_{\rF,\, n+r,\,s}(\{\kappa_m\})}\otimes\ket{\tilde{P}^{(h\to 0)}_{\rF,\, n-r,\,s}(\{\kappa_m\})}. 
\label{Schmidt_F}
\ee
Here, 
\bea
\ket{\tilde{P}^{(0\to h)}_{\rF,\, n+r,\,s}(\{\kappa_m\})} & = & 
\frac{1}{\sqrt{\tilde{N}_{\rF,\,n+r,\,s}^{(0\to h)}}}\,\sum_{w\in\tilde{P}^{(0\to h)}_{\rF,\, n+r,\,s}(\{\kappa_m\})}\ket{w}, \\
\ket{\tilde{P}^{(h\to 0)}_{\rF,\, n-r,\,s}(\{\kappa_m\})} & = &
\frac{1}{\sqrt{\tilde{N}_{\rF,\,n-r,\,s}^{(h\to 0)}}}\,\sum_{w\in\tilde{P}^{(h\to 0)}_{\rF,\, n-r,\,s}(\{\kappa_m\})}\ket{w}, 
\eea
and 
\be
p^{(h)}_{\rF,\,n+r,n-r,\,s}=\frac{\tilde{N}_{\rF,\,n+r,\,s}^{(0\to h)} \tilde{N}_{\rF,\,n-r,\,s}^{(h\to 0)}}{N_{\rF,\, 2n, \,s}}
=s^{-h}\frac{N^{(h)}_{\rF,\,n+r}N^{(h)}_{\rF,\,n-r}}{N_{\rF,\, 2n}}.
\label{pnr_F}
\ee

From the density matrix of the ground state $\rho_S=\ket{P_{\,\rF,\,2n,\,s}}\bra{P_{\,\rF,\,2n,\,s}}$ with (\ref{Schmidt_F}), 
the reduced density matrix is obtained as 
\be
\rho_A = \Tr_B\,\rho_S = \sum_{h=0}^{n-|r|}\sum_{\kappa_1=1}^s\cdots \sum_{\kappa_h=1}^s p^{(h)}_{\rF,\,n+r,n-r,\,s}\,
\ket{\tilde{P}^{(0\to h)}_{\rF,\, n+r,\,s}(\{\kappa_m\})}\bra{\tilde{P}^{(0\to h)}_{\rF,\, n+r,\,s}(\{\kappa_m\})}, 
\label{rhoA_F}
\ee
where we used the orthonormal property: 
\be
\left\langle\tilde{P}^{(h\to 0)}_{\rF,\, n-r,\,s}(\{\kappa_m\})\right.\ket{\tilde{P}^{(h'\to 0)}_{\rF,\, n-r,\,s}(\{\kappa'_m\})} 
=\delta_{h,h'}\delta_{\kappa_1,\kappa'_1}\cdots \delta_{\kappa_h,\kappa'_h}.
\ee
Since $\rho_A$ is a diagonal form, the EE (\ref{vNS}) is recast as
\be
S_{\rF,\,A} = -\sum_{h=0}^{n-|r|}s^h\,p^{(h)}_{\rF,\, n+r,n-r,\,s}\ln p^{(h)}_{\rF,\, n+r,n-r,\,s}.
\label{vNS_F} 
\ee
Note that $p^{(h)}_{\rF,\, n+r,n-r,\,s}$ does not depend on $\kappa_1, \cdots, \kappa_h$ 
and the sums $\sum_{\kappa_1=1}^s\cdots \sum_{\kappa_h=1}^s$ yield the factor $s^h$. 

We plug (\ref{N2n_F}) and (\ref{Nhn_F}) into (\ref{pnr_F}) and evaluate its asymptotic behavior as~\cite{fredkin} (see also \cite{sk_renyi,sk_renyi2})
\bea
 p^{(h)}_{\rF,\,n+r,n-r,\,s} &\sim & s^{-h} \frac{1+(-1)^{n+r+h}}{2}\,\frac{8}{\sqrt{\pi}}\,\left(\frac{n}{(n+r)(n-r)}\right)^{3/2}(h+1)^2\,e^{-\frac{n\,(h+1)^2}{(n+r)(n-r)}}
 \nn \\
& & \,\times \left[1+O(n^{-1})\right].
\label{pnr_asym_F2}
\eea
By converting the sum in (\ref{vNS_F}) to an integral, we compute the EE as 
\bea
S_{\rF,\,A} & = & (2\ln s)\sqrt{\frac{(n+r)(n-r)}{\pi n}} + \frac12\ln \frac{(n+r)(n-r)}{n} +\frac12\ln \frac{\pi}{4} + \gamma-\frac12-\ln s 
\nn \\
& & + (\mbox{terms vanishing as $n\to \infty$}).
\label{vNS_Ff}
\eea
The leading term scales as a square-root of $n$ and significantly violates the area law in spite of local interactions. 
Note that this originates from the following part of (\ref{vNS_F}):
\be
-\sum_{h=0}^{n-|r|}s^h\,p^{(h)}_{\rF,\, n+r,n-r,\,s}\ln \left(s^{-h}\right)=
\frac{\ln s}{N_{\rF,\,2n, \,s}}\sum_{h=0}^{n-|r|} h s^h\, \tilde{N}_{\rF,\,n+r,\,s}^{(0\to h)} \tilde{N}_{\rF,\,n-r,\,s}^{(h\to 0)}, 
\label{vNS_F_leading}
\ee 
which implies the factor $s^{-h}$ in $p^{(h)}_{\rF,\, n+r,n-r,\,s}$ is crucial to get the $\sqrt{n}$-scaling.

\section{Motzkin spin chain}
\label{sec:motzkin}
\setcounter{equation}{0}
The Motzkin spin chain~\cite{motzkin} has additional spin degrees of freedom (we call zero-spin) at each site compared with the Fredkin spin chain. 
In total, there are up- and down-spin states with color $k\in\{1,2,\cdots,s\}$ and the zero-spin state at the site $i$, denoted by $\ket{u^k_i}$, $\ket{d^k_i}$ and $\ket{0_i}$, respectively.  
The Hamiltonian of the Motzkin spin chain of length $2n$ is given in the form of the sum of projection operators:
\bea
H_{\rM, s} & = & \sum_{j=1}^{2n-1}\sum_{k=1}^s \left\{\ket{U^k_{j, j+1}}\bra{U^k_{j, j+1}}+
\ket{D^k_{j, j+1}}\bra{D^k_{j, j+1}} + \ket{F^k_{j, j+1}}\bra{F^k_{j, j+1}}\right\} \nn \\
& & +\sum_{j=1}^{2n-1}\sum_{k\neq\ell}\ket{u^k_j, d^\ell_{j+1}}\bra{u^k_j, d^\ell_{j+1}} 
+ \sum_{k=1}^s\left\{\ket{d^k_1}\bra{d^k_1} + \ket{u^k_{2n}}\bra{u^k_{2n}}\right\},
\label{HM}
\eea
where
\bea
\ket{U^k_{j, j+1}} & = & \frac{1}{\sqrt{2}}\left(\ket{0_j, u^k_{j+1}}
-\ket{u^k_j, 0_{j+1}}\right), \\
\ket{D^k_{j, j+1}} & = & \frac{1}{\sqrt{2}}\left(\ket{0_j, d^k_{j+1}}
-\ket{d^k_j, 0_{j+1}}\right),\\
\ket{F^k_{j, j+1}} & = & \frac{1}{\sqrt{2}}\left(\ket{0_j, 0_{j+1}}
-\ket{u^k_j, d^k_{j+1}}\right),
\eea
and the interactions are among nearest neighbors. 

The Hamiltonian has a unique ground state at zero-energy. 
We can repeat the same identification of the spins and steps as before with the additional zero-spin corresponding to the arrow $(1,0)$ (flat-step).  
For colorless case ($s=1$), the ground state is expressed by the equal-weight superposition of length-$2n$ Motzkin walks, 
which are random walks consisting of up-, down- and flat-steps, 
starting at the origin, ending at $(2n, 0)$ and not allowing paths to enter $y<0$ region. 
For $s$-color case ($s>1$), the color assigned to each up-step 
should be matched with that of the subsequent down-step at the same height, which is the same as in the Fredkin spin chain. 

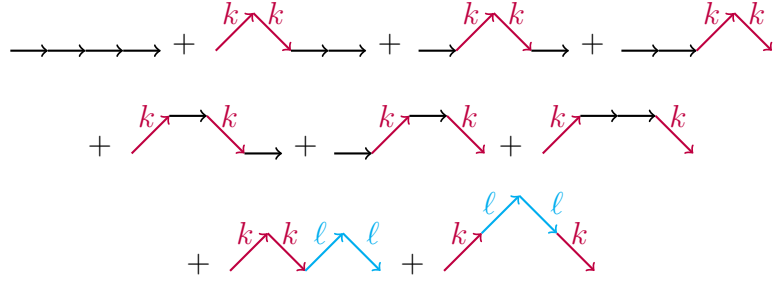
\begin{figure}[H]
\centering
\captionsetup{width=.8\linewidth}
\begin{tikzpicture}
\draw[->, thick] (0,0)--(0.5,0);
\draw[->, thick] (0.5,0)--(1,0);
\draw[->, thick] (1,0)--(1.5,0);
\draw[->, thick] (1.5,0)--(2,0);
\end{tikzpicture} 
$+ \,$ 
\begin{tikzpicture}
\draw[purple, ->, thick] (0,0)--(0.5,0.5);
\draw[purple, ->, thick] (0.5,0.5)--(1,0);
\draw[->, thick] (1,0)--(1.5,0);
\draw[->, thick] (1.5,0)--(2,0);
\node (k) at (0.2,0.5) {\textcolor{purple}{$k$}};
\node (k) at (0.8,0.5) {\textcolor{purple}{$k$}};
\end{tikzpicture} 
$+ \,$
\begin{tikzpicture}
\draw[->, thick] (0,0)--(0.5,0);
\draw[purple,->, thick] (0.5,0)--(1,0.5);
\draw[purple, ->, thick] (1,0.5)--(1.5,0);
\draw[->, thick] (1.5,0)--(2,0);
\node (k) at (0.7,0.5) {\textcolor{purple}{$k$}};
\node (k) at (1.3,0.5) {\textcolor{purple}{$k$}};
\end{tikzpicture} 
$+ \,$
\begin{tikzpicture}
\draw[->, thick] (0,0)--(0.5,0);
\draw[->, thick] (0.5,0)--(1,0);
\draw[purple, ->, thick] (1,0)--(1.5,0.5);
\draw[purple, ->, thick] (1.5,0.5)--(2,0);
\node (k) at (1.2,0.5) {\textcolor{purple}{$k$}};
\node (k) at (1.8,0.5) {\textcolor{purple}{$k$}};
\end{tikzpicture} 
\\
\vspace{5mm}
$+ \,$
\begin{tikzpicture}
\draw[purple, ->, thick] (0,0)--(0.5,0.5);
\draw[->, thick] (0.5,0.5)--(1,0.5);
\draw[purple, ->, thick] (1,0.5)--(1.5,0);
\draw[->, thick] (1.5,0)--(2,0);
\node (k) at (0.2,0.5) {\textcolor{purple}{$k$}};
\node (k) at (1.3,0.5) {\textcolor{purple}{$k$}};
\end{tikzpicture} 
$ + \,$
\begin{tikzpicture}
\draw[->, thick] (0,0)--(0.5,0);
\draw[purple, ->, thick] (0.5,0)--(1,0.5);
\draw[->, thick] (1,0.5)--(1.5,0.5);
\draw[purple, ->, thick] (1.5,0.5)--(2,0);
\node (k) at (0.7,0.5) {\textcolor{purple}{$k$}};
\node (k) at (1.8,0.5) {\textcolor{purple}{$k$}};
\end{tikzpicture} 
$+ \,$ 
\begin{tikzpicture}
\draw[purple, ->, thick] (0,0)--(0.5,0.5);
\draw[->, thick] (0.5,0.5)--(1,0.5);
\draw[->, thick] (1,0.5)--(1.5,0.5);
\draw[purple, ->, thick] (1.5,0.5)--(2,0);
\node (k) at (0.2,0.5) {\textcolor{purple}{$k$}};
\node (k) at (1.8,0.5) {\textcolor{purple}{$k$}};
\end{tikzpicture} 
\\
\vspace{3mm}
$+\,$
\begin{tikzpicture}
\draw[purple, ->, thick] (0,0)--(0.5,0.5);
\draw[purple, ->, thick] (0.5,0.5)--(1,0);
\draw[cyan, ->, thick] (1,0)--(1.5,0.5);
\draw[cyan, ->, thick] (1.5,0.5)--(2,0);
\node (k) at (0.2,0.5) {\textcolor{purple}{$k$}};
\node (k) at (0.8,0.5) {\textcolor{purple}{$k$}};
\node (k') at (1.2,0.5) {\textcolor{cyan}{$\ell$}};
\node (k') at (1.9,0.5) {\textcolor{cyan}{$\ell$}};
\end{tikzpicture} 
$+ \,$
\begin{tikzpicture}
\draw[purple, ->, thick] (0,0)--(0.5,0.5);
\draw[cyan, ->, thick] (0.5,0.5)--(1,1);
\draw[cyan, ->, thick] (1,1)--(1.5,0.5);
\draw[purple, ->, thick] (1.5,0.5)--(2,0);
\node (k) at (0.2,0.5) {\textcolor{purple}{$k$}};
\node (k') at (0.6,0.9) {\textcolor{cyan}{$\ell$}};
\node (k') at (1.5,0.9) {\textcolor{cyan}{$\ell$}};
\node (k) at (1.8,0.5) {\textcolor{purple}{$k$}};
\end{tikzpicture}
\caption{Colored Motzkin walks corresponding to (\ref{PM4_Motzkin}). Up- and down-steps with the same color are matched.}
\label{fig:PM4}       
\end{figure}

The ground state is expressed as 
\be
\ket{P_{\rM,\,2n,\,s}}=\frac{1}{\sqrt{N_{\rM,\,2n,\,s}}}\sum_{w\in P_{\rM,\,2n,\,s}}\ket{w},
\ee
where $N_{\rM,\,2n,\,s}$ in the normalization factor is the number of the length-$2n$ colored Motzkin walks given by 
\be
N_{\rM,\,2n,\,s}=\sum_{\rho=0}^{n}\binomi{2n}{2\rho}s^{n-\rho}\,N_{\rF,\,2n-2\rho},
\label{N2n_M}
\ee
where $2\rho$ stands for the number of the flat-steps. 
For example, $2n=4$ case reads
\begin{align}
\ket{P_{\rM,\, 4,\, s}} &= \,\frac{1}{\sqrt{1+6s+2s^2}}\nn \\
& \times\Biggl[  \ket{0_1, 0_2, 0_3, 0_4}  
+ \sum_{k=1}^s \left\{\ket{u^k_1, d^k_2, 0_3, 0_4} + \ket{0_1, u^k_2, d^k_3, 0_4} +\ket{0_1, 0_2, u^k_3, d^k_4} \right.\nn \\
& \hspace{42mm}\left.+\ket{u^k_1, 0_2, d^k_3, 0_4} + \ket{0_1, u^k_2, 0_3, d^k_4} + \ket{u^k_1, 0_2, 0_3, d^k_4}\right\} \nn \\
& \hspace{32mm}+\sum_{k,\ell=1}^s \left\{\ket{u^k_1, d^k_2, u^\ell_3, d^\ell_4} + \ket{u^k_1, u^\ell_2, d^\ell_3,d^k_4}\right\}\Biggr],
\label{PM4_Motzkin}
\end{align}
which corresponds to the colored Motzkin walks in Fig.~\ref{fig:PM4}.
%

\subsection{EE of the ground state}
Computing in the same manner as in the previous section, we obtain 
\be
S_{\rM,\,A} = -\sum_{h=0}^{n-|r|}s^h\,p^{(h)}_{\rM,\, n+r,n-r,\,s}\ln p^{(h)}_{\rM,\, n+r,n-r,\,s}
\label{vNS_M} 
\ee
with 
\bea
p^{(h)}_{\rM,\,n+r,n-r,\,s} & = & s^{-h}\,\frac{N_{\rM,\,n+r,\,s}^{(0\to h)}N_{\rM,\,n-r,\,s}^{(h\to 0)}}{N_{\rM,\, 2n, \,s}},
\label{pnr_M}
\\
N_{\rM,\, n\pm r,\, s}^{(0\to h)} = N_{\rM,\,n\pm r,\,s}^{(h\to 0)} 
& = &\sum_{\rho=0}^{n\pm r-h}\binomi{n\pm r}{\rho} N^{(h)}_{\rF,\, n\pm r-\rho}\,\, s^{\frac{n\pm r-\rho+h}{2}} . 
\label{N_M}
\eea

The asymptotic form of $p^{(h)}_{\rM,\,n+r,n-r,\,s}$ is evaluated as~\cite{motzkin} (see also \cite{sk_renyi,sk_renyi2}) 
\bea
p^{(h)}_{\rM,\, n+r,n-r,\,s} & \sim & s^{-h}\sqrt{\frac{2}{\pi\sigma^3}}\,\left(\frac{n}{(n+r)(n-r)}\right)^{3/2}(h+1)^2\,e^{-\frac{1}{2\sigma } \frac{n\,(h+1)^2}{(n+r)(n-r)}} 
\nn \\
& & \times \left[1+O(n^{-1})\right]
\label{pnr_asym_M2}
\eea
with $\sigma\equiv \frac{\sqrt{s}}{2\sqrt{s}+1}$. 
Finally, we end up with 
\bea
S_{\rM,\,A} & = & (2\ln s)\sqrt{\frac{2\sigma}{\pi}\frac{(n+r)(n-r)}{n}} + \frac12\ln \frac{(n+r)(n-r)}{n} +\frac12\ln (2\pi\sigma) + \gamma-\frac12-\ln s \nn \\
& & + 
(\mbox{terms vanishing as $n\to \infty$}).
\label{vNS_Ms}
\eea
Again, the leading term grows as a square-root of $n$ that is beyond the logarithmic violation of the area law usually seen in critical systems, 
although interactions of the Hamiltonian (\ref{HM}) are local. 
This behavior originates from the following part of (\ref{vNS_M}):
\be
-\sum_{h=0}^{n-|r|}s^h\,p^{(h)}_{\rM,\, n+r,n-r,\,s}\ln \left(s^{-h}\right)=
\frac{\ln s}{N_{\rM,\,2n, s}}\sum_{h=0}^{n-|r|} h s^h\, \tilde{N}_{\rM,\,n+r,\,s}^{(0\to h)} \tilde{N}_{\rM,\,n-r,\,s}^{(h\to 0)}.
\label{vNS_M_leading}
\ee

\section{Large-$N$ matrix models}
\label{sec:MM}
\setcounter{equation}{0}
In this section, we consider large-$N$ matrix models which reproduce the $\sqrt{n}$-scaling of the EEs in (\ref{vNS_Ff}) and (\ref{vNS_Ms}).  

\subsection{Case of Fredkin spin chain}
\begin{figure}[H]
\centering
\captionsetup{width=.8\linewidth}
\begin{tikzpicture}
\draw (0,0)--(4.2,0);
\fill (0.2,0) circle [radius=2pt];
\foreach \x in {1,2,3,4} \draw (\x,0.2)--(\x,-0.2);
\node (1) at (1,-0.5) {1};
\node (2) at (2,-0.5) {2};
\node (3) at (3,-0.5) {3};
\node (4) at (4,-0.5) {4};
\draw [purple,ultra thick] (1.53,0) arc (0:180:0.53cm);
\draw [purple,ultra thick] (1.48,0) arc (0:180:0.48cm);
\draw [cyan,ultra thick] (3.53,0) arc (0:180:0.53cm);
\draw [cyan,ultra thick] (3.48,0) arc (0:180:0.48cm);
\node (k) at (1,0.76) {\textcolor{purple}{$k$}};
\node (ell) at (3,0.76) {\textcolor{cyan}{$\ell$}};
\node (plus) at (5,0) {$+$};
\end{tikzpicture}
$\quad$
\begin{tikzpicture}
\draw (0,0)--(4.2,0);
\fill (0.2,0) circle [radius=2pt];
\foreach \x in {1,2,3,4} \draw (\x,0.2)--(\x,-0.2);
\node (1) at (1,-0.5) {1};
\node (2) at (2,-0.5) {2};
\node (3) at (3,-0.5) {3};
\node (4) at (4,-0.5) {4};
\draw [purple,ultra thick] (3.53,0) arc (0:180:1.53cm);
\draw [purple,ultra thick] (3.48,0) arc (0:180:1.48cm);
\draw [cyan,ultra thick] (2.53,0) arc (0:180:0.53cm);
\draw [cyan,ultra thick] (2.48,0) arc (0:180:0.48cm);
\node (k) at (2,1.25) {\textcolor{purple}{$k$}};
\node (ell) at (2,0.76) {\textcolor{cyan}{$\ell$}};
\end{tikzpicture}
\caption{Rainbow diagrams in $2n=4$ case.
In each of these two, the black dot is a marked point, and the left and right edges are identified to make a loop. Semi-circles represent contractions by the propagator. 
The first and second terms corresponds to the first and second terms in Fig.~\ref{fig:PF4}, respectively. 
A semi-circle corresponds to a color-matched up- and down-spins. 
} 
\label{fig:rainbow4}
\end{figure}
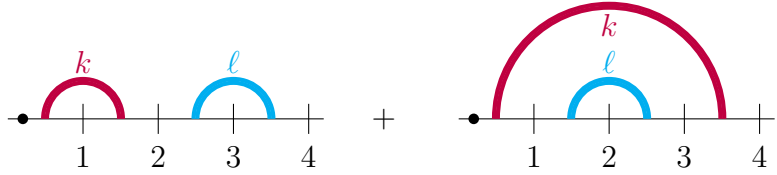

Let us start with one-to-one correspondence between colored Dyck walks and rainbow diagrams in the Gaussian matrix model of $N\times N$ hermitian matrices $M_f$ ($f=1,2,\cdots,s$):
\bea
& & S_{G_1} =  N\sum_{f=1}^s\tr \left(\frac12M_f^2\right), \nn \\
& & Z_{G_1}=\int\left(\prod_{f=1}^sd^{N^2}M_f\right)\,e^{-S_{G_1}}, \qquad \vev{\cdot}_{G_1}=\frac{1}{Z_{G_1}}\int\left(\prod_{f=1}^sd^{N^2}M_f\right)\,e^{-S_{G_1}}\,(\cdot).
\label{ZG_cM}
\eea
The one-point function 
\be
\vev{\frac{1}{N}\tr\left(\cM^{2n}\right)}_{G_1} \qquad \mbox{with} \qquad \cM\equiv \sum_{f=1}^sM_f
\ee
is expressed as the sum of rainbow diagrams in the large-$N$ limit. 
The operator $\frac{1}{N}\tr\left(\cM^{2n}\right)$ makes a length-$2n$ loop with a marked point. Feynman graphs for the one-point function are drawn by all possible pairwise contractions of $2n$ $\cM$s by the propagator 
\be
\vev{\left(M_f\right)_{ij}\left(M_{f'}\right)_{k\ell}}_{G_1}=\frac{1}{N}\delta_{ff'}\delta_{i\ell}\delta_{jk}.
\label{prop_cM}
\ee
In the large-$N$ limit, there remain only planar diagrams among these that are called as rainbow diagrams~\cite{rainbow}. For example, Fig.~\ref{fig:rainbow4} shows rainbow diagrams for $2n=4$ case. 
We can see that each semi-circle of the rainbow diagrams has one-to-one correspondence to a color-matched up- and down-spin pairs. Thus, 
\be
\lim_{N\to \infty}  \vev{\frac{1}{N}\tr\left(\cM^{2n}\right)}_{G_1}=N_{\rF,\,2n,\,s}.
\label{F-cM}
\ee
%

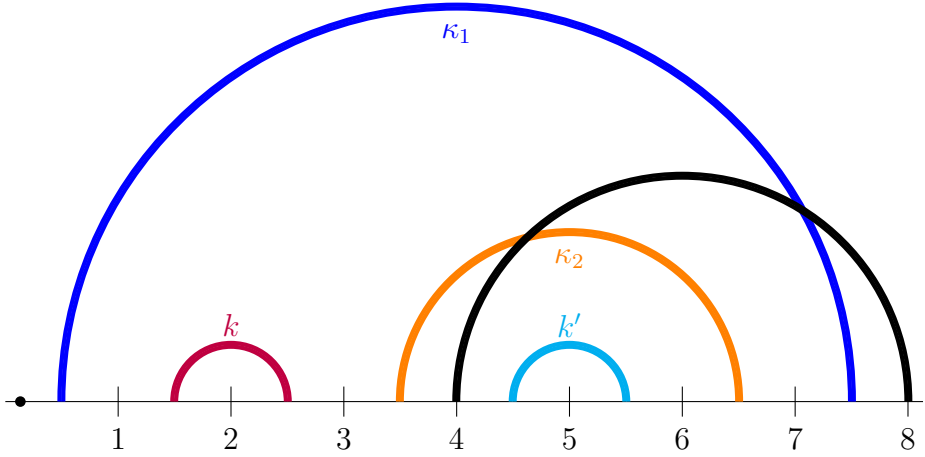
\begin{figure}[H]
\centering
\captionsetup{width=.8\linewidth}
\begin{tikzpicture}
\draw (0,0)--(12.2,0);
\fill (0.2,0) circle [radius=2pt];
\foreach \x in {1.5,3,4.5,6,7.5,9,10.5,12} \draw (\x,0.2)--(\x,-0.2);
\node (1) at (1.5,-0.5) {1};
\node (2) at (3,-0.5) {2};
\node (3) at (4.5,-0.5) {3};
\node (4) at (6,-0.5) {4};
\node (5) at (7.5,-0.5) {5};
\node (6) at (9,-0.5) {6};
\node (7) at (10.5,-0.5) {7};
\node (8) at (12,-0.5) {8}; 
\draw[purple,ultra thick] (3.78,0) arc (0:180:0.78cm);
\draw[purple,ultra thick] (3.73,0) arc (0:180:0.73cm);
\draw[cyan,ultra thick] (8.28,0) arc (0:180:0.78cm);
\draw[cyan,ultra thick] (8.23,0) arc (0:180:0.73cm);
\draw[orange,ultra thick] (9.78,0) arc (0:180:2.28cm);
\draw[orange,ultra thick] (9.73,0) arc (0:180:2.23cm);
\draw[blue,ultra thick] (11.28,0) arc (0:180:5.28cm);
\draw[blue,ultra thick] (11.23,0) arc (0:180:5.23cm);
\draw[ultra thick] (12.03,0) arc (0:180:3.03cm);
\draw[ultra thick] (11.98,0) arc (0:180:2.98cm);
\node (k) at (3,1.02) {\textcolor{purple}{$k$}};
\node (k') at (7.5,1.02) {\textcolor{cyan}{$k'$}};
\node (kappa1) at (6,4.9) {\textcolor{blue}{$\kappa_1$}};
\node (kappa2) at (7.5,1.9) {\textcolor{orange}{$\kappa_2$}};
\end{tikzpicture}
\caption{The diagram corresponding to the divided path in Fig.~\ref{fig:PF8}. The black dot is a marked point, and the left and right edges are identified. 
Each colored semi-circle corresponds to a color-matched up-and down-spins, and the black semi-circle to the boundary between 
the subsystems $A$ and $B$. The two crosses of the black and colored lines represent the operators $\tr \left(M_{\kappa_1} XM_{\kappa_1} X\right)$ and 
$\tr \left(M_{\kappa_2} XM_{\kappa_2} X\right)$.  
}
\label{fig:rainbow8}
\end{figure}

Next, for the remaining part of (\ref{vNS_F_leading}), we consider the operator 
\be
\frac{1}{N}\tr \left(\cM^{n+r}X\cM^{n-r}X\right)
\label{operator_cM_X}
\ee 
by introducing another $N\times N$ hermitian matrix $X$. 
$\cM^{n+r}$ and $\cM^{n-r}$ represent spins in the subsystems $A$ and $B$ respectively, and $X$ borders of the subsystems. We can see that the connected correlation function 
\be
\vev{\frac{1}{N}\tr \left(\cM^{n+r}X\cM^{n-r}X\right)\, \frac{1}{h!}\left\{\frac{N}{2}\sum_{f=1}^s\tr \left(M_f XM_f X\right)\right\}^h}_{G_2,\,{\rm connected}}
\label{connected}
\ee
evaluated by the Gaussian action $S_{G_2}\equiv N\tr \left(\sum_{f=1}^s \frac12M_f^2 + \frac12X^2\right)$ 
reproduces $s^h\, \tilde{N}_{\rF,\,n+r,\,s}^{(0\to h)} \tilde{N}_{\rF,\,n-r,\,s}^{(h\to 0)}$ 
as $N\to \infty$, as far as we ignore diagrams including any contraction by $M_f$-propagators among $\left\{\frac{N}{2}\sum_{f=1}^s\tr \left(M_f XM_f X\right)\right\}^h$. 
In Fig.~\ref{fig:rainbow8}, we show the diagram corresponding to the divided path in Fig.~\ref{fig:PF8}.  
%

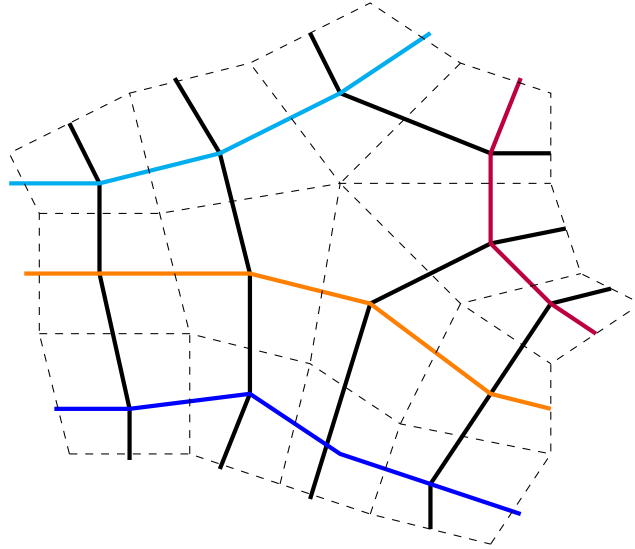
\begin{figure}[H]
\centering
\captionsetup{width=.8\linewidth}
\begin{tikzpicture}[scale=0.4]
\draw[ultra thick] (4,2.8)--(4,4.5)--(3,9)--(3,12)--(2,14);
\draw[ultra thick] (7,2.5)--(8,5)--(8,9)--(7,13)--(5.5,15.5);
\draw[ultra thick] (10,1.5)--(12,8)--(16,10)--(18.5,10.5);
\draw[ultra thick] (14,0.5)--(14,2)--(16,5)--(18,8)--(20,8.5);
\draw[ultra thick] (10,17)--(11,15)--(16,13)--(18,13);
\draw[cyan,ultra thick] (0,12)--(3,12)--(7,13)--(11,15)--(14,17);
\draw[orange,ultra thick] (0.5,9)--(3,9)--(8,9)--(12,8)--(16,5)--(18,4.5);
\draw[blue,ultra thick] (1.5,4.5)--(4,4.5)--(8,5)--(11,3)--(14,2)--(17,1);
\draw[purple,ultra thick] (17,15.5)--(16,13)--(16,10)--(18,8)--(19.5,7);
\draw[dashed] (2,3)--(1,7)--(1,11)--(0,13);
\draw[dashed] (6,3)--(6,7)--(5,11)--(4,15);
\draw[dashed] (9,2)--(10,6)--(11,12)--(8,16);
\draw[dashed] (12,1)--(13,4)--(15,8)--(19,9);
\draw[dashed] (16,0)--(18,3)--(18,6)--(21,8)--(19,9)--(18,12)--(18,15)--(15,16)--(12,18)--(8,16)--(4,15)--(0,13);
\draw[dashed] (1,11)--(5,11)--(11,12)--(18,12);
\draw[dashed] (1,7)--(6,7)--(10,6)--(13,4)--(18,3);
\draw[dashed] (2,3)--(6,3)--(9,2)--(12,1)--(16,0);
\draw[dashed] (15,16)--(11,12)--(15,8)--(18,6);
\end{tikzpicture}
\caption{A part of a typical loop diagram and its dual. Black and colored lines represent the $X$- and $M_f$- propagators, respectively. Their crosses are the four-point vertices. 
Each loop in the diagram consists of an alternating combination of black and colored lines. 
Dashed lines indicate the dual diagram, which can be interpreted as a randomly quadrangulated surface. 
 }
\label{fig:loop_graph}       
\end{figure}

This observation naturally leads to a matrix model action with so-called $ABAB$-type interactions:
\be
S=N\tr\left[\sum_{f=1}^s \frac12 M_f^2 + \frac12 X^2 +\frac{g}{2}\sum_{f=1}^s M_f X M_f X\right],
\label{S_cM_X}
\ee
under which the one-point function of (\ref{operator_cM_X}) evaluated by tree and planar diagrams gives
\be
\left.\lim_{N\to \infty}\vev{\frac{1}{N}\tr \left(\cM^{n+r}X\cM^{n-r}X\right)}\right|_{\rm tree}=\sum_{h=0}^\infty (-g)^h s^h \, \tilde{N}_{\rF,\,n+r,\,s}^{(0\to h)} \tilde{N}_{\rF,\,n-r,\,s}^{(h\to 0)}. 
\ee
Here, tree diagrams do not include any loop composed solely by internal lines (propagators) and the four-point vertices $\tr\left(M_f X M_f X\right)$ ($f=1,2,\cdots, s$). 
Note that the vertices should be located along the line connecting two $X$s in (\ref{operator_cM_X}). Otherwise, the diagram will not be a tree or planar graph. 
Finally we find 
\be
\left.\lim_{g\to -1}\lim_{N\to \infty}g\frac{\partial}{\partial g}\vev{\frac{1}{N}\tr \left(\cM^{n+r}X\cM^{n-r}X\right)}\right|_{{\rm tree}} 
= \sum_{h=0}^\infty h s^h \, \tilde{N}_{\rF,\,n+r,\,s}^{(0\to h)} \tilde{N}_{\rF,\,n-r,\,s}^{(h\to 0)}.
\label{EEF_num_cM}
\ee
The sum in (\ref{vNS_F_leading}) is given by evaluating the connected two-point function of the operators (\ref{operator_cM_X}) 
and $\frac12\sum_{f=1}^s \tr \left(M_f X M_f X\right)$ at the tree and planar level. 
Although at the tree level, the Feynman diagrams look like skeletons and would not allow an interpretation as a smooth random surface, diagrams including loop effects are expected to allow. 
A part of a typical loop diagram of the matrix model is drawn in Fig.~\ref{fig:loop_graph}.   
That can give a generalization of the usual EE that includes fluctuating bulk geometry.   
%

We can also express (\ref{F-cM}) as 
\be
\left.\lim_{N\to \infty}\vev{\frac{1}{N}\tr \left(\cM^{2n}\right)}\right|_{\rm tree}=N_{\rF,\,2n,\,s}.
\label{F-cM2}
\ee 

\subsection{Simpler matrix model}
We note that the same conclusion as in the above is reached by a simpler matrix model:
\bea
& & S_s\equiv N\tr\left[\frac{1}{2s}M^2+\frac12+\frac{g}{2s}MXMX\right],
\nn \\
& & Z_s=\int d^{N^2}M\,d^{N^2}X\,e^{-S_s}, \qquad \vev{\cdot}_s=\frac{1}{Z_s}\int d^{N^2}M\,d^{N^2}X\,e^{-S_s}\,(\cdot).
\label{Zs_M_X} 
\eea
Whereas $M_f$ in (\ref{ZG_cM}) or (\ref{S_cM_X}) represents contribution by a single color $f$, the $N\times N$ hermitian matrix $M$ in (\ref{Zs_M_X}) carries 
contribution of the whole $s$ colors.  

In this model, we obtain analogous results to (\ref{F-cM2}) and (\ref{EEF_num_cM}) as
\be 
\left.\lim_{N\to \infty}\vev{\frac{1}{N}\tr (M^{2n})}_s\right|_{\rm tree}=N_{\rF,\,2n,\,s}
\label{F-M}
\ee 
and 
\be
\left.\lim_{g\to -1}\lim_{N\to \infty}g\frac{\partial}{\partial g}\vev{\frac{1}{N}\tr \left(M^{n+r}XM^{n-r}X\right)}_s\right|_{{\rm tree}} 
= \sum_{h=0}^\infty h s^h \, \tilde{N}_{\rF,\,n+r,\,s}^{(0\to h)} \tilde{N}_{\rF,\,n-r,\,s}^{(h\to 0)}.
\label{EEF_num_M}
\ee

\subsection{Case of Motzkin spin chain}
For the Motzkin spin chain, we can use the same matrix models as in the above with simple modification to operators. 

Let us consider the operator $\frac{1}{N}\tr\left(\cM+F\right)^{2n}$, where $F$ is some $N\times N$ matrix. In the binomial expansion of $\left(\cM+F\right)^{2n}$, 
we can see that $\cM$ represents the up- and down-spins with color as before, and $F$ corresponds to the zero-spin. 
Note that terms including the odd numbers of $\cM$ vanish in its one-point function due to the $Z_2$-symmetry under $M_f\to -M_f$ for each $f$. 
As far as concerning the EE, namely the number of the Motzkin paths, we may set $F$ to the identity matrix in the computation. Thus, 
\be
\lim_{N\to \infty} \vev{\frac{1}{N}\tr\left[\left(\cM+\id_N\right)^{2n}\right]}_{G_1} = N_{\rM,\,2n,\,s}
\label{M-cM}
\ee 
for the Gaussian matrix model (\ref{ZG_cM}), or 
\be
\left.\lim_{N\to \infty} \vev{\frac{1}{N}\tr\left[\left(\cM+\id_N\right)^{2n}\right]}\right|_{\rm tree} = N_{\rM,\,2n,\,s} 
\label{M-cM2}
\ee
for the matrix model given by the action (\ref{S_cM_X}). 

For the division to the two subsystems, we obtain 
\be
\left.\lim_{g\to -1}\lim_{N\to \infty}g\frac{\partial}{\partial g}\vev{\frac{1}{N}\tr \left[\left(\cM+\id_N\right)^{n+r}X\left(\cM+\id_N\right)^{n-r}X\right]}\right|_{{\rm tree}} 
= \sum_{h=0}^\infty h s^h \, \tilde{N}_{\rM,\,n+r,\,s}^{(0\to h)} \tilde{N}_{\rM,\,n-r,\,s}^{(h\to 0)}.
\label{EEM_num_cM}
\ee

We can repeat the same argument for the simpler matrix model (\ref{Zs_M_X}) with the result
\be 
\left.\lim_{N\to \infty}\vev{\frac{1}{N}\tr \left[(M+\id_N)^{2n}\right]}_s\right|_{\rm tree}=N_{\rM,\,2n,\,s}
\label{M-M}
\ee 
and 
\be
\left.\lim_{g\to -1}\lim_{N\to \infty}g\frac{\partial}{\partial g}\vev{\frac{1}{N}\tr \left[\left(M+\id_N\right)^{n+r}X\left(M+\id_N\right)^{n-r}X\right]}_s\right|_{{\rm tree}} 
= \sum_{h=0}^\infty h s^h \, \tilde{N}_{\rM,\,n+r,\,s}^{(0\to h)} \tilde{N}_{\rM,\,n-r,\,s}^{(h\to 0)}.
\label{EEM_num_M}
\ee

\section{Matrix model solution and extended EE}
\label{sec:MM_sol}
\setcounter{equation}{0}
The matrix models (\ref{S_cM_X}) and (\ref{Zs_M_X}) have so-called $ABAB$ interactions, to which we cannot apply the standard method to 
reduce eigenvalue integrals~\cite{IZ,mehta}. 
However, exact solutions have been obtained by applying a technique to solve $O(n)$ model on a random surface~\cite{CK} or by using character expansion~\cite{KZJ}.~\footnote{
Case of a general potential for $X$ is treated in \cite{CK}, and the action considered in~\cite{KZJ} contains a common quartic self-interaction term of each matrix.} 
In what follows, we focus on the simpler matrix model (\ref{Zs_M_X}) or equivalently the model defined by the action: 
\be
S_s'=N\tr\left[\frac12M'^2+\frac12X^2+\frac{g}{2}M'XM'X\right]\quad \mbox{with} \quad M'=\frac{1}{\sqrt{s}}M. 
\label{S'_M_X}
\ee
In what follows, $\vev{\cdot}_s'$ denotes an expectation value evaluated by $S_s'$. 

As discussed in appendix~\ref{app:MM_sol}, this model becomes critical at $g=g_c\equiv -\frac29$. 
The coupling constant $g$ counts the number of the vertices in Feynman diagrams. As $g$ approaches to $g_c$, diagrams which consist of large number of the vertices dominantly contribute. 
Since each vertex appearing in the diagram represented by a plaquette in its dual graph, Feynman diagrams are interpreted as randomly quadrangulated surfaces by the plaquettes. 
Namely, this defines a lattice model for two-dimensional quantum gravity~\cite{david2}. 
In this section, we always consider the large-$N$ limit, which corresponds to extracting surfaces of planar topology, and suppress the symbol ``$\lim_{N\to \infty}$'' for notational simplicity.   
We introduce a lattice spacing $a$ (length of the edges of the plaquette), and take the continuum limit of the lattice model as $a\to 0$, $g\to g_c$ 
with physical quantities (for example, area) fixed finite. 

As shown in the end of appendix~\ref{app:MM_sol}, the model (\ref{S'_M_X}) can describe pure quantum gravity in the two-dimensional bulk 
with the string susceptibility exponent $\gamma_{\rm str}=-1/2$~\cite{KPZ,david,DK}. 
However, it exhibits a different scaling relation between boundary length and bulk area 
\be
\vev{\mbox{(boundary length)}}\sim\vev{\mbox{(area)}}^{3/4}
\label{boundary_area}
\ee 
from what we have seen in the pure gravity ($\vev{\mbox{(boundary length)}}\sim\vev{\mbox{(area)}}^{1/2}$). 

\subsection{Case of Fredkin spin chain}
We compute an analog of the leading term of the EE for the Fredkin spin chain defined by 
\be
S_{\rF,\,A}^{\rm grav.}\equiv \frac{g\frac{\partial}{\partial g}\vev{\frac{1}{N}\tr \left(M^{n+r}XM^{n-r}X\right)}_s}{\vev{\frac{1}{N}\tr (M^{2n})}_s}\,\ln s,
\label{SFAg}
\ee
which is an extension of (\ref{vNS_F_leading}) with (\ref{F-M}) and (\ref{EEF_num_M}), thereby removing the restriction to the tree diagrams. 
(\ref{SFAg}) is evaluated around the critical point. 
 
\subsubsection{$\vev{\frac{1}{N}\tr (M^{2n})}_s$}
First, we evaluate large-$n$ behavior of $\vev{\frac{1}{N}\tr (M^{2n})}_s$. 
From (\ref{app:omega_hat}), $z'\omega(z')=z'\vev{\frac{1}{N}\tr\frac{1}{z'-M'}}_s'$ behaves near the critical point as 
\be
z'\omega(z') = -4\cdot 2^{1/4}\,a t^{1/8}\sqrt{\zeta'+t^{3/4}}+\cdots
\label{omega_hat}
\ee
with 
\be 
g=g_c\left(1-\frac32\,a^2t\right), \qquad z'=\frac{3}{\sqrt{2}}\left(1+\frac12a^{3/2}\zeta'\right).
\label{g_z'_scaling}
\ee
$t$ and $\zeta'$ stand for bulk and boundary cosmological constants in the continuum theory that control 
area and boundary length, respectively. In order to consider contribution from large area and large boundary length, we ignore terms analytic in $t$ or $\zeta'$. 
Here and below, the ellipsis means such irrelevant terms. 
Let us see large-order behavior of the expansion of 
\be
a^{3/4}\sqrt{\zeta'+t^{3/4}}=\sqrt{1-\frac{\lambda_*^2}{z'^2}}\,\times [1+O(a^{3/2})]
\label{sqrt_exp_F}
\ee 
by $\frac{\lambda_*^2}{z'^2}$, where $\lambda_*$ is given in (\ref{app:lamb*}). 
For example, from 
\be
\binomi{\nu}{n}\sim(-1)^{n+1}\frac{\sin(\pi\nu)}{\pi}\Gamma(\nu+1) \,n^{-\nu-1}\left[1+O(n^{-1})\right] \qquad (\nu\notin \bZ),
\label{binomial}
\ee
we obtain 
\be
z'\omega(z')=2\cdot 2^{1/4}a^{1/4}t^{1/8}\sum_n\frac{1}{\sqrt{\pi}}\frac{1}{n^{3/2}} \left(\frac{\lambda_*^2}{z'^2}\right)^n+\cdots. 
\label{z'omega}
\ee
Since $z\vev{\frac{1}{N}\tr\frac{1}{z-M}}_s$ in the model (\ref{Zs_M_X}) is nothing but $z'\omega(z')$ with $z=\sqrt{s}\,z'$, we read off large-$n$ behavior of $\vev{\frac{1}{N}\tr\left(M^{2n}\right)}_s$ as
\be
\vev{\frac{1}{N}\tr\left(M^{2n}\right)}_s\sim 2\cdot 2^{1/4}a^{1/4}t^{1/8}\frac{1}{\sqrt{\pi}}\frac{1}{n^{3/2}}\left(\frac92\,s\right)^n.
\label{EEF_den_G}
\ee

This should be compared with behavior of (\ref{N2n_F}) or (\ref{F-M}):
\be
\left.\vev{\frac{1}{N}\tr (M^{2n})}_s\right|_{\rm tree}=N_{\rF,\,2n,\,s}\sim\frac{1}{\sqrt{\pi}}\frac{1}{n^{3/2}}\,(4s)^n.
\ee
The power of $n$ is common. 
Effects of fluctuating bulk surface are found in $t$-dependence in the overall factor and $\frac92\,s=4s\cdot \frac98$.  

\subsubsection{$g\frac{\partial}{\partial g}\vev{\frac{1}{N}\tr \left(M^{n+r}XM^{n-r}X\right)}_s$}
Next, we obtain asymptotic behavior of $g\frac{\partial}{\partial g}\vev{\frac{1}{N}\tr \left(M^{n+r}XM^{n-r}X\right)}_s$. 

From (\ref{app:omega(2)e2}), $z_1'z_2'\omega^{(2)}_e(z_1',z_2')=z_1'^2z_2'^2\vev{\frac{1}{N}\tr\left(\frac{1}{z_1'^2-M'^2}X\frac{1}{z_2'^2-M'^2}X\right)}_s'$ behaves as
\be
z_1'z_2\omega^{(2)}_e(z_1',z_2')=-16\sqrt{2}\,a^{1/2}t^{1/4}\,\frac{\sqrt{\zeta_1'+t^{3/4}}\sqrt{\zeta'_2+t^{3/4}}}{\zeta_1'+\zeta_2'} + \cdots,
\ee
where $z_i'=\frac{3}{\sqrt{2}}\left(1+\frac12a^{3/2}\zeta_i'\right)$ ($i=1,2$). After taking $t$-derivative, 
we have 
\bea
\lefteqn{g\frac{\partial}{\partial g}z_1'z_2'\omega^{(2)}_e(z_1',z_2') = \frac{8\sqrt{2}}{3}\,a^{-3/2}\left[\frac32\frac{1}{\sqrt{\zeta_1'+t^{3/4}}\sqrt{\zeta_2'+t^{3/4}}}\right. } \nn \\
& &  \left.+t^{-3/4}\frac{\sqrt{\zeta_1+t^{3/4}}\sqrt{\zeta_2'+t^{3/4}}}{\zeta_1'+\zeta_2'}+\frac{3t^{3/4}}{(\zeta_1'+\zeta_2')\sqrt{\zeta_1'+t^{3/4}}\sqrt{\zeta_2'+t^{3/4}}}\right]+\cdots.
\label{dg_omega(2)e}
\eea
We will see that these three terms provide different scalings. Use of (\ref{sqrt_exp_F}), 
$
\frac{1}{\zeta_1'+\zeta_2'}=a^{3/2}\sum_{L=0}^\infty \left(\frac{9/2}{z_1'z_2'}\right)^{2L}
$
and 
\be
\sum_{L=0}^{\min\{k,\,\ell\}}\binomi{1/2}{k-L}\binomi{1/2}{\ell-L}=-(-1)^{k+\ell}\frac{2(k+\ell)+1}{4(k-\ell)^2-1}\frac{1}{2^{2(k+\ell)}}\frac{(2k)!(2\ell)!}{\left(k!\,\ell!\right)^2}
\label{sum_binomials}
\ee
leads to the large-order series
\begin{align}
g\frac{\partial}{\partial g}z_1'z_2'\omega^{(2)}_e(z_1',z_2') = &\frac{8\sqrt{2}}{3\pi}\sum_{k,\,\ell} \left\{\frac{3}{2\sqrt{k\ell}}-a^{3/2}t^{-3/4}\frac{2(k+\ell)}{4(k-\ell)^2-1}\frac{1}{\sqrt{k\ell}} \right. \nn \\
&  \hspace{7mm}\left.+3a^{3/2}t^{3/4}\sum_{L\geq 0}\frac{1}{\sqrt{(k-L)(\ell-L)}}\right\}\left(\frac92\,s\right)^{k+\ell} \frac{1}{z_1^{2k}}\frac{1}{z_2^{2\ell}}
\label{series_F}
\end{align}
with $z_i=\sqrt{s}\,z_i'$ ($i=1,2$). 
Since the boundary length scales as $a^{-3/2}$ from (\ref{g_z'_scaling}), we define the length in the continuum: 
\be
b\equiv a^{3/2}n, \qquad u\equiv a^{3/2}r.
\label{b_u}
\ee
Then, the sum of $L$ in (\ref{series_F}) is evaluated for $k=\frac{n+r}{2}$, $\ell=\frac{n-r}{2}$ as 
\be
\sum_{L\geq 0}\frac{2}{\sqrt{(n+r-2L)(n-r-2L)}}=2\ln\frac{\sqrt{b+u}+\sqrt{b-u}}{\sqrt{2|u|}}
\label{sum_L}
\ee
for $u\Slash{\sim} 0$. Reading off the coefficient of $1/(z_1^{n+r}z_2^{n-r})$ we find 
\begin{align}
g\frac{\partial}{\partial g}\vev{\frac{1}{N}\tr \left(M^{n+r}XM^{n-r}X\right)}_s = &  \frac{8\sqrt{2}}{3\pi}\,a^{3/2}\left\{\frac{3}{\sqrt{(b+u)(b-u)}}-t^{-3/4}\frac{b}{u^2\sqrt{(b+u)(b-u)}} \right. \nn \\
& \hspace{14mm} \left. +6t^{3/4}\ln\frac{\sqrt{b+u}+\sqrt{b-u}}{\sqrt{2|u|}}\right\}\left(\frac92\,s\right)^{2n}. 
\label{EEF_num_G}
\end{align}

\subsubsection{Extended EE with fluctuating bulk geometry}
Now we take the ratio of (\ref{EEF_num_G}) and (\ref{EEF_den_G}), and obtain the extended EE (\ref{SFAg}) as
\bea
S_{\rF,\, A}^{\rm grav.}& = & \frac{4\cdot 2^{1/4}}{3\sqrt{\pi}}\,a^{-1} t^{-1/8}\left\{\frac{3b^{3/2}}{\sqrt{(b+u)(b-u)}}-t^{-3/4}\frac{b^{5/2}}{u^2\sqrt{(b+u)(b-u)}} \right. \nn \\
& & \hspace{27mm} \left. +6t^{3/4}b^{3/2}\ln\frac{\sqrt{b+u}+\sqrt{b-u}}{\sqrt{2|u|}}\right\} \ln s.
\label{vNS_F_G}
\eea
From the derivation, this expression is valid unless $u$ is around the origin or $\pm b$. 
For $|u|$ being of the same order as $b$ (typically $|u|=\frac12b, \frac13b$, etc.) or smaller, the third term becomes dominant scaling as $b^{3/2}$ or $b^{3/2}\ln b$, which implies that the bulk quantum gravity greatly enhances the square-root scaling (\ref{vNS_Ff}). 
It would be intuitively understandable because it seems that fishnet diagrams describing a random surface provide stronger correlations than rainbow diagrams at the tree level. 
Although $r$- or $u$-dependence is different, the first term of (\ref{vNS_F_G}) reproduces the square-root scaling. 
The first and third terms come from the first and third terms in (\ref{dg_omega(2)e}), respectively. Difference between the two, namely the factor $1/(\zeta_1'+\zeta_2')$, is crucial to the enhancement. 
It seems natural because the factor cannot be factorized as a product of a function of $\zeta_1'$ and a function of $\zeta_2'$, representing the entanglement of the subsystems $A$ and $B$.   
Whereas (\ref{vNS_Ff}) has the maximum at $r=0$ (separation at the middle), ({\ref{vNS_F_G}) grows as $|u|$ increases when $|u|\ll b$. This suggests that fluctuating geometry 
provides quite different behavior from (\ref{vNS_Ff}) on the dependence of the separation.

\subsection{Case of Motzkin spin chain}
For case of the Motzkin spin chain, in view of (\ref{vNS_M_leading}), (\ref{M-M}) and (\ref{EEM_num_M}), we compute the extended EE
\be
S_{\rM,\,A}^{\rm grav.}\equiv \frac{g\frac{\partial}{\partial g}\vev{\frac{1}{N}\tr \left[(M+\id_N)^{n+r}X(M+\id_N)^{n-r}X\right]}_s}{\vev{\frac{1}{N}\tr \left[(M+\id_N)^{2n}\right]}_s}\,\ln s
\label{SMAg}
\ee
around the critical point. 

First, a generating function of $\vev{\frac{1}{N}\tr \left[(M+\id_N)^{2n}\right]}_s$ is related to $\omega(z')$ as 
\be
z\vev{\frac{1}{N}\tr\frac{1}{z-(M+\id_N)}}_s=\frac{\sqrt{s}\,z'+1}{\sqrt{s}}\,\omega(z')
\label{resolventM_omega}
\ee
with
\be
z'=\frac{z-1}{\sqrt{s}}.
\label{z'_z_M}
\ee
In (\ref{omega_hat}), we consider large-order behavior in the expansion of 
\be
a^{3/4}\sqrt{\zeta'+t^{3/4}}=\frac{2^{3/4}}{\sqrt{3}}\sqrt{z'-\lambda_*}=\left(\frac{3\sqrt{s}+\sqrt{2}}{\sqrt{2s}}\right)^{1/2}\sqrt{1-\frac{\sqrt{s}\lambda_*+1}{z}},
\label{sqrt_exp_M}
\ee
and obtain 
\be
\vev{\frac{1}{N}\tr \left[(M+\id_N)^{2n}\right]}_s\sim 2^{1/4}a^{1/4}t^{1/8}\frac{1}{\sqrt{\pi}\,\sigma_G^{3/2}}\,\frac{1}{n^{3/2}}\left(\sqrt{s}\lambda_{*0}+1\right)^{2n}
\label{EEM_den_G}
\ee
with
$\sigma_G\equiv \frac{3\sqrt{s}}{3\sqrt{s}+\sqrt{2}}$ and $\lambda_{*0}\equiv \lim_{a\to 0}\lambda_*=\frac{3}{\sqrt{2}}$. 
As comparison, (\ref{N_M}) or (\ref{M-M}) behaves as 
\be
N_{\rM,\,2n,\,s}\sim \frac{1}{2\sqrt{\pi}\,\sigma^{3/2}}\frac{1}{n^{3/2}}\,(2\sqrt{s}+1)^{2n} \qquad \left(\sigma=\frac{\sqrt{s}}{2\sqrt{s}+1}\right). 
\ee
We recognize similar effects to the case of the Fredkin model.   

Next, for $g\frac{\partial}{\partial g}\vev{\frac{1}{N}\tr \left[\left(M+\id_N\right)^{n+r}X\left(M+\id_N\right)^{n-r}X\right]}_s$, we do similar computation to the Fredkin case. 
In $\omega^{(2)}(z_1',z_2')$ given in (\ref{app:omega(2)2}), we use (\ref{sqrt_exp_M}), $\frac{1}{\zeta_1'+\zeta_2'}= \frac{\sigma_G}{2}\,a^{3/2} \sum_{L=0}^\infty\left(\frac{(\sqrt{s}\lambda_{*0}+1)^2}{z_1z_2}\right)^L$ 
and (\ref{sum_binomials}) to obtain 
\begin{align}
g\frac{\partial}{\partial g}z_1'z_2'\omega^{(2)}(z_1',z_2') = & \frac{16\sqrt{2}}{3\pi} \sum_{k,\,\ell}\left\{\frac34\sigma_G\frac{1}{\sqrt{k\ell}} -a^{-3/2}t^{-3/4}\frac{2(k+\ell)}{4(k-\ell)^2-1}\frac{1}{\sqrt{k\ell}} \right. \nn \\
 & \hspace{17mm}\left. +\frac34\,a^{3/2}t^{3/4}\sigma_G^2\sum_{L\geq 0}\frac{1}{\sqrt{(k-L)(\ell-L)}}\right\} \frac{\left(\sqrt{s}\lambda_{*0}+1\right)^{k+\ell}}{z_1^kz_2^{\ell}}.
\end{align}
Converting variables as (\ref{b_u}), we find  
\bea
\lefteqn{g\frac{\partial}{\partial g}\vev{\frac{1}{N}\tr \left[\left(M+\id_N\right)^{n+r}X\left(M+\id_N\right)^{n-r}X\right]}_s} \nn \\
& & = \frac{16\sqrt{2}}{3\pi}\,a^{3/2}\left\{\frac{3}{4\sigma_G}\frac{1}{\sqrt{(b+u)(b-u)}} -\frac{1}{4\sigma_G^2}\,t^{-3/4} \frac{b}{u^2\sqrt{(b+u)(b-u)}} \right.\nn \\
 & & \hspace{22mm}\left. +\frac32\,t^{3/4} \ln\frac{\sqrt{b+u}+\sqrt{b-u}}{\sqrt{2|u|}}\right\}\left(\sqrt{s}\lambda_{*0}+1\right)^{2n}
\label{EEM_num_G}
\eea 
for $u\Slash{\sim} 0$.  

 Combining (\ref{EEM_den_G}) and (\ref{EEM_num_G}), we end up with the extended EE including fluctuating bulk geometry as 
 \bea
 S_{\rM,\,A}^{\rm grav.} & = & \frac{16\cdot 2^{1/4}}{3\sqrt{\pi}}\,a^{-1}t^{-1/8}\left\{\frac{3\sigma_G^{1/2}}{4}\frac{b^{3/2}}{\sqrt{(b+u)(b-u)}} 
 -\frac{1}{4\sigma_G^{1/2}}\,t^{-3/4} \frac{b^{5/2}}{u^2\sqrt{(b+u)(b-u)}} \right.\nn \\
 & & \hspace{22mm}\left. +\frac{3\sigma_G^{3/2}}{2}\,t^{3/4} \ln\frac{\sqrt{b+u}+\sqrt{b-u}}{\sqrt{2|u|}}\right\}\ln s,
\label{vNS_M_G} 
\eea
which has essentially the same structure as in the Fredkin case (\ref{vNS_F_G}).

\section{Discussions}
\label{sec:summary}
\setcounter{equation}{0}
In this paper, we introduce large-$N$ matrix models, which reproduce the leading terms of the EE in highly entangled spin chains in their Feynman diagrams at the tree and planar level. 
By using the exact solution in one of such models, we compute analogous quantity to the EE including the full loop effects in the large-$N$ limit. 
Although diagrams look like skeletons at the tree level, a two-dimensional random surface emerges in the bulk by including the loop effects. 
The effects greatly increase the entanglement from the square-root scaling to the scaling of the power $3/2$ (with logarithmic correction), and make change the dependence of the separation: 
the entanglement grows as the difference of the length of the subsystems $A$ and $B$ increases, as far as the difference is small. 
An intuitive explanation to the former is that fishnet diagrams describing a random surface provide much more correlation than the skeletons.   
It will be interesting to understand a physical meaning of the latter property. 

Since the $(s+1)$-matrix model of $M_f$ ($f=1,\cdots, s$) and $X$ can express more details of spin configurations compared to the two-matrix model of $M$ and $X$, 
it will be important to analyze the $(s+1)$-matrix model and gain deeper insights into the system. For $s\leq 2$ the exact solution is found in \cite{CK}. 
In that case, it will be nice if any technique is developed to obtain 
the relevant one-point functions $\vev{\frac{1}{N}\tr\left(\cM^{2n}\right)}$ and $\vev{\frac{1}{N}\tr\left(\cM^{n+r}X\cM^{n-r}X\right)}$ with $\cM=\sum_{f=1}^sM_f$.  
      
It will also be worth doing analogous investigation for other entanglement measures like R\'enyi entanglement entropy~\cite{sk_renyi,sk_renyi2} and mutual information~\cite{dellanna}, 
and extending to deformed Motzkin/Fredkin spin chains 
in which the EEs grow linearly~\cite{zhang_k,salberger_etal,zhang_ak}.  

In the matrix models, the operators corresponding to spin configurations are regarded as one-dimensional objects, whereas their Feynman diagrams naturally generate two-dimensional surfaces.  
It seems intriguing to investigate the matrix models from the viewpoint of holographic (random) tensor networks~\cite{ludwig,klich1,klich2}.

\section*{Acknowledgements}
The author thanks to Branko Dragovich for suggesting publication of this paper. 
This research was supported by the Institute for Basic Science in Korea (IBS-R018-D1). 

\appendix
\section{Exact solution to matrix model}
\label{app:MM_sol}
\setcounter{equation}{0}
In this appendix, we present the exact solution to the large-$N$ matrix model defined by the action 
\be
S_s'=N\tr\left[\frac12M'^2+\frac12X^2+\frac{g}{2}M'XM'X\right] ,
\label{app:S'_M_X}
\ee
which is equivalent to (\ref{Zs_M_X}) with $M'= \frac{1}{\sqrt{s}}M$. 

\subsection{Character expansion}
As discussed in~\cite{KZJ}, we consider the expansion by $GL(N)$ characters $\chi_R$:  
\be
e^{-\frac{Ng}{2}\tr\left(M'XM'X\right)}=\sum_Rd_R\zeta_R(Ng)\chi_R(M'X),
\label{app:ch_exp}
\ee 
where $R$ in the sum runs over polynomial irreducible representations labelled by shifted highest weights $h_i=N-i+m_i$ ($i=1,2 \cdots,N$). 
$\{m_i\}$ and $\{h_i\}$ are sets of nonnegative integers satisfying  
\be
m_1\geq m_2 \geq \cdots \geq m_N\geq 0 \qquad \mbox{and}\qquad h_1>h_2>\cdots >h_N\geq 0.
\ee
In terms of $\{h_i\}$, $\chi_R(A)=\det_{j,\,k}\left(a^{h_j}_k\right)/\triangle(a)$ with $a_i$ being eigenvalues of $A$ and $\triangle(a)$ the Van der Monde determinant 
$\triangle(a)\equiv \det_{j,\,k}\left(a^{N-j}_k\right)=\prod_{j<k}(a_j-a_k)$.  
$d_R$ denotes the dimension of the representation $R$: $d_R=\chi_R(\id_N)=\prod_{j<k}\frac{h_j-h_k}{k-j}$. 
The expansion coefficient $\zeta_R(Ng)$ is given by integrals over $U(N)$ matrices: 
\be
\zeta_R(Ng)=\frac{1}{d_R}\int[dU]\,\chi_R(U^\dagger)\,e^{-\frac{Ng}{2}\tr(U^2)}
\label{app:zeta_R}
\ee 
with $[dU]$ the $U(N)$ Haar measure normalized as $\int [dU]=1$. Let us consider case of $N$ even. Then, it can be seen that each representation $\{h_j\}$ contributing to the expansion (\ref{app:ch_exp}) consists of 
$N/2$ even integers and $N/2$ odd integers. For the representation specified by even integers $\{h^e\}\equiv\{h_{j_1},\cdots, h_{j_{N/2}}\}$ and 
odd integers $\{h^o\}\equiv \{h_{j_{N/2 +1}},\cdots,h_{j_N}\}$, 
(\ref{app:zeta_R}) is given as 
\be
\zeta_R(Ng)=\frac{1}{d_R}\left(-\frac{Ng}{2}\right)^{\frac12|m|}(-1)^{\frac{N(N+2)}{8}}\epsilon_{j_1\cdots j_N}
\frac{\triangle\left(h^e\right)\triangle\left(h^o\right)}{2^{\frac{N(N-2)}{4}}\prod_{j=1}^N\left\lfloor\frac{h_j}{2}\right\rfloor!},
\ee
where $|m|\equiv \sum_{i=1}^Nm_i=\sum_{i=1}^Nh_i-\frac{N(N-1)}{2}$, 
$\epsilon_{j_1\cdots j_N}$ is an $N$-th rank totally antisymmetric tensor normalized by $\epsilon_{1\cdots N}=1$, and $\lfloor x \rfloor$ denotes the greatest integer not exceeding $x$. 

Now by using the formula 
\be
\int [dU]\,\chi_R(AUBU^\dagger)=\frac{1}{d_R}\chi_R(A)\chi_R(B),
\ee
the partition function reduces to eigenvalue integrals as
\be
Z_s'=\int d^{N^2}M'\,d^{N^2}X\,e^{-S_s'}=C_N\sum_{\{h\}}\left(-\frac{Ng}{2}\right)^{\frac12|h|-\frac{N(N-1)}{2}}\,c_{\{h\}}R_{\{h\}}(g)^2
\label{app:Zs'}
\ee
with $C_N$ being a constant depending only on $N$, 
\bea
c_{\{h\}} & \equiv & \frac{1}{\prod_{i=1}^N \left\lfloor\frac{h_j}{2}\right\rfloor!\,\prod_{h_j\in\{h^e\},\,h_k\in\{h^o\}}(h_j-h_k)}, \\
R_{\{h\}}(g) &\equiv & \int \left(\prod_{i=1}^N d\lambda_i\right)\,\triangle(\lambda)\,\det_{j,\,k}\left(\lambda^{h_j}_k\right)\,e^{-N\sum_{i=1}^N\frac12\lambda_i^2}.
\label{app:Rhg}
\eea

\subsection{Large-$N$ analysis of $R_{\{h\}}(g)$}
A large-$N$ saddle point for (\ref{app:Rhg}) is given by the equation 
\be
-\lambda_i^2+\frac{1}{N}\lambda_i\sum_{k(\neq i)}\frac{1}{\lambda_i-\lambda_k}+\frac{1}{N}\lambda_i\frac{\partial}{\partial\lambda_i}\ln\det_{j,\,k}\left(\lambda^{h_j}_k\right)=0.
\label{app:spe_R}
\ee
Following the standard analysis in \cite{BIPZ}, we introduce the resolvent~\footnote{Correlation functions evaluated by (\ref{app:Zs'}) with (\ref{app:S'_M_X}) are denoted by $\vev{\cdot}_s'$.}  
\be
\omega(\lambda)=\frac{1}{N}\sum_{i=1}^N\frac{1}{\lambda-\lambda_i}=\vev{\frac{1}{N}\tr \frac{1}{\lambda-M'}}_s'=\vev{\frac{1}{N}\tr \frac{1}{\lambda-X}}_s'
\label{app:omega}
\ee
for $\lambda\in \bC$, and assume that the eigenvalue density $\rho_\lambda(x)\equiv \frac{1}{N}\sum_{i=1}^N\delta(x-\lambda_i)$ becomes continuous with some support $[-\lambda_*,\,\lambda_*]$ 
as $N\to \infty$. Then. 
\be
\omega(\lambda)=\int^{\lambda_*}_{-\lambda_*}dy\,\frac{\rho_\lambda(y)}{\lambda-y}
\label{app:omega2}
\ee 
is analytic in $\lambda\in\bC$ except the cut $[-\lambda_*,\,\lambda_*]$, and 
$\omega(\lambda)=\frac{1}{\lambda} + O(\lambda^{-3})$ as $\lambda\to \infty$.
$\frac{1}{N}\sum_{k(\neq i)}\frac{1}{\lambda_i-\lambda_k}$ in (\ref{app:spe_R}) is given by 
$\Slash{\omega}(x) \equiv \frac12\left(\omega(x+i0)+\omega(x-i0)\right)$ for $x\in [-\lambda_*,\,\lambda_*]$. We also introduce a holomorphic function $h(\lambda)$ such that 
$h(\lambda)$ has the same cut as $\omega(\lambda)$ and 
\be
\Slash{h}(x)\equiv \frac12\left(h(x+i0)+h(x-i0)\right) =\left.\frac{1}{N}\lambda_i\frac{\partial}{\partial\lambda_i}\ln \det_{j,\,k}\left(\lambda^{h_j}_k\right)\right|_{\lambda_i=x}
\ee
for $x\in [-\lambda_*,\,\lambda_*]$. The saddle point equation (\ref{app:spe_R}) can be expressed as 
\be
-x^2+x\Slash{\omega}(x)+\Slash{h}(x)=0\qquad (x\in  [-\lambda_*,\,\lambda_*]).
\label{app:spe_R2}
\ee
Note that $\chi_{\{h\}}\left({\rm diag}(\lambda_1,\cdots,\lambda_N)\right)=\det_{j,\,k}\left(\lambda^{h_j}_k\right)/\triangle(\lambda)$ is a polynomial of $\lambda_1,\cdots,\lambda_N$ and can be extended to 
$\lambda_1,\cdots,\lambda_N\in \bC$. By taking $\lambda_i$-derivative of the logarithm of the character, we see that $\Slash{h}(\lambda_i)-\lambda_i\Slash{\omega}(\lambda_i)$ is 
extendible to the whole complex plane. 
Thus, 
\be
h(\lambda)-\lambda\omega(\lambda)=f(\lambda) \quad \mbox{for}\quad \lambda\in \bC
\label{app:h_omega}
\ee
with $f(\lambda)$ having no cut~\cite{ZJ}. From (\ref{app:spe_R2}) and (\ref{app:h_omega}), we obtain
\be
-\lambda^2+\lambda\omega(\lambda)+h^\dagger(\lambda)=0  \quad \mbox{for}\quad \lambda\in \bC,
\label{app:spe_R3}
\ee
where  $h^\dagger$ denotes $h$ on the second Riemann sheet.  For large $\lambda$, (\ref{app:spe_R3}) yields 
\be
h^\dagger(\lambda)=\lambda^2-1-\sum_{m=1}^\infty\frac{1}{\lambda^{2m}}\vev{\frac{1}{N}\tr \left(M'^{2m}\right)}_s'.
\label{app:expand_h}
\ee
Note that $\vev{\frac{1}{N}\tr\left(M'^m\right)}_s'=0$ for $m$ odd from the $Z_2$ symmetry under $M'\to -M'$. 

\subsection{Large-$N$ analysis for highest weights}
In order to consider saddle point equations for highest weights $\{h_i\}$, we define 
\be
H(h)=\frac{1}{N}\sum_{i=1}^N\frac{1}{h-h_i'}
\label{app:H}
\ee
with $h_i'\equiv \frac{1}{N}h_i$. Putting $x=\frac{i}{N}$, we assume that $h(x)=h_i'$ becomes a continuous function in large-$N$ limit. 
A typical distribution of the highest weights is that $m_k=m_{k+1}=\cdots =m_N=0$ for some $k$ and the rest nonzero. Setting 
$\alpha=h_1'$ and $\beta=h_k'$, we then see that $h(x)$ decreases along the slope of $-x$ for $\frac{k}{N}<x<1$, and more rapidly decreases for $0<x<\frac{k}{N}$.   
This implies that the highest weight density 
$\rho(h)=-\frac{\partial}{\partial h}x(h)$ saturates as $\rho(h)=1$ for $0<h<\beta$ and $0<\rho(h)<1$ for $\beta<h<\alpha$. 
(\ref{app:H}) can be written as 
\be
H(h) =  \int^\alpha_0dy\,\frac{\rho(y)}{h-y}=\ln\frac{h}{h-\beta}+\int^\alpha_\beta dy\, \frac{\rho(y)}{h-y},
\label{app:H2}
\ee
which is analytic except the cut $[0,\alpha]$, $H(h) = \frac{1}{h} + O(h^{-2})$ as $h\to \infty$, and 
\be
H(h\pm i0)=\Slash{H}(h) \mp i\pi\rho(h) \quad (h\in[0,\alpha])
\ee
with $\Slash{H}(h)=\frac12\left(H(h+i0)+H(h-i0)\right)=\int_0^\alpha dy\,\rho(y)\,{\rm P}\frac{1}{h-y}$. 
Next, let us introduce a function $L(h)$ such that $L(h)$ has the same cut $[0,\alpha]$ as $H(h)$ and 
\be
\Slash{L}(y)\equiv\frac12\left(L(y+i0)+L(y-i0)\right)= \left.\frac{2}{N}\frac{\partial}{\partial h'_i}\ln \det_{j,\,k}\left(\lambda_k^{Nh'_j}\right) \right|_{h'_i=y}.
\ee
Note that $\lambda(h)\equiv \exp\left(\frac12L(h)\right)$ and $h(\lambda)$ are functional inverses of each other as mutli-valued functions as discussed in~\cite{ZJ}.   
The inversion of (\ref{app:expand_h}) is iteratively done as 
\be
\lambda(h)^2=h^2+1+\frac{1}{h}\vev{\frac{1}{N}\tr \left(M'^2\right)}_s'+O\left(y^{-2}\right).
\label{app:lam-h_exp}
\ee

The large-$N$ saddle point equation of (\ref{app:Zs'}) with respect to $h_i'$ reads~\footnote{As usually found in large-$N$ limit~\cite{BIPZ}, 
models are well-defined for some negative region of coupling constants. Here we also consider the case of $g<0$.} 
\be  
2\Slash{L}(h)-\Slash{H}(h)
=\ln\frac{h}{-g} \quad \mbox{for} \quad h\in[\beta,\alpha]. 
\label{app:spe_h}
\ee
In terms of the analytic function 
\be
D(h)\equiv 2L(h)-H(h)+\ln(h-\beta)-3\ln h,
\label{app:Dh}
\ee
which asymptotically behaves as
\be
D(h) = \frac{1-\beta}{h}+O(h^{-2}),
\label{app:Dh_asymp}
\ee
(\ref{app:spe_h}) is recast as 
\be
\Slash{D}(h) =\ln\frac{h-\beta}{(-g)h^2}\quad \mbox{for}\quad h\in[\beta,\alpha].
\label{app:spe_h2}
\ee  
Its solution is given by 
\be
D(h)=\lim_{\epsilon\to +0}\sqrt{(h-\alpha)(h-\beta)}\oint_C\frac{ds}{2\pi i}\frac{\ln \frac{s-\beta+\epsilon}{(-g)s^2}}{(h-s)\sqrt{(s-\alpha)(s-\beta)}},
\ee
where the integration contour $C$ encloses only the square-root cut $[\beta,\alpha]$ but not the other singularities. The integral is evaluated by inflating the contour as 
\be
D(h) =\ln\frac{(\alpha-\beta)(h-\beta)\left(\sqrt{h-\alpha}+\sqrt{h-\beta}\right)^2}{(-g)\left(\sqrt{\alpha(h-\beta)}+\sqrt{\beta(h-\alpha)}\right)^4}.
\label{app:sol_D}  
\ee
(\ref{app:Dh_asymp}) requires that $X\equiv\left(\frac{\sqrt{\alpha}+\sqrt{\beta}}{2}\right)^2$ satisfies 
\be
X-3g^2X^3=1,
\label{app:X_eq}
\ee
the suitable solution of which reads 
\be
X=-\frac{2}{3g}\,{\rm Im}\left[\sqrt{1-\frac{81}{4}g^2}-i\frac92g\right]^{1/3}
\label{app:X_sol}
\ee
behaving as $X=1+3g^2+O(g^4)$ for $g\sim 0$.  
The solution determines
$\alpha$ and $\beta$ as 
\be
\alpha=X(1-gX)^2,\qquad \beta=X(1+gX)^2.
\ee

The critical point of $g$ is given by a singular point of (\ref{app:X_sol}) nearest from the origin: $g_c=-\frac29$. 
Near the critical point, expansion in $\Delta\equiv \frac23\frac{g-g_c}{-g_c}$ leads to
\be
X=\frac32\left[1-\Delta^{1/2}+O(\Delta)\right]  , \qquad \alpha = \frac83\left[1-\frac32\Delta^{1/2}+O(\Delta)\right], \qquad \beta=\frac23\left[1+O(\Delta)\right].
\ee

\subsection{Solution of $\omega(\lambda)$}
It can be directly seen that (\ref{app:sol_D}) is regular except the cut $[\beta,\alpha]$. Together with (\ref{app:Dh_asymp}), we identify $D(h)$ with $H(h)-\ln\frac{h}{h-\beta}$. 
Then, 
\begin{align}
H(h) = & \ln\frac{(\alpha-\beta)h\left(\sqrt{h-\alpha}+\sqrt{h-\beta}\right)^2}{(-g)\left(\sqrt{\alpha(h-\beta)}+\sqrt{\beta(h-\alpha)}\right)^4} ,
\label{app:sol_H}\\
\rho(y) = & \frac{2}{\pi}\left\{2\,{\rm arg}\left(\sqrt{\alpha(y-\beta)}+i\sqrt{\beta(\alpha-y)}\right)-{\rm arg}\left(\sqrt{y-\beta} +i \sqrt{\alpha-y}\right)\right\} 
\end{align}
for $y\in[\beta,\alpha]$.
Also, we find 
\be
\lambda(h)^2=e^{L(h)}=h\,e^{H(h)},
\ee
leading to 
\be
\lambda(h)=\frac{\sqrt{\alpha-\beta}\,h\left(\sqrt{h-\alpha}+\sqrt{h-\beta}\right)}{\sqrt{-g}\left(\sqrt{\alpha(h-\beta)}+\sqrt{\beta(h-\alpha)}\right)^2}
= \frac{\left(h+\sqrt{\alpha\beta}+\sqrt{(h-\alpha)(h-\beta)}\right)^2}{2h\left(\sqrt{h-\alpha}+\sqrt{h-\beta}\right)}.
\label{app:sol_lamb}
\ee

The critical point $h=h_*$ of (\ref{app:sol_lamb}) satisfying $\lambda'(h_*)=0$ is
\be
h_*=2\sqrt{\alpha\beta}=\frac83\left[1-\frac34\Delta^{1/2}+O(\Delta)\right].
\label{app:h*}
\ee
The functional inversion of $\lambda(h)$ around the critical point gives
\be
h(\lambda)=h_* \pm\frac{8\cdot 2^{1/4}}{3}\Delta^{1/8}\left[1+O(\Delta^{1/2})\right] \sqrt{\lambda-\lambda_*}+O(\lambda-\lambda_*)
\label{app:sol_h}  
\ee
with 
\be
\lambda_*=\frac{3}{\sqrt{2}}\left[1-\frac12\Delta^{3/4}+O(\Delta)\right]. 
\label{app:lamb*}
\ee
Since $h^\dagger (\lambda)$ is given by (\ref{app:sol_h}) with $\pm$ replaced by $\mp$. From (\ref{app:spe_R3}), 
\bea
\omega(\lambda) & = & \frac{1}{\lambda}\left(\lambda^2-h^\dagger(\lambda)\right) \nn \\
 & = & \lambda_*-\frac{h_*}{\lambda_*} \pm \frac{8\cdot 2^{1/4}}{3}\Delta^{1/8}\left[1+O(\Delta^{1/2})\right] \sqrt{\frac{\lambda-\lambda_*}{\lambda_*}}+O(\lambda-\lambda_*).
\label{app:sol_omega}  
\eea 
Choices of the branches are fixed by $\rho_\lambda(x)\geq 0$. We should take the ``$-$" branch in (\ref{app:sol_omega}) and (\ref{app:sol_h}). 

\subsection{Continuum limit}
Since Feynman diagrams of the matrix model (\ref{app:S'_M_X}) are interpreted as randomly quadrangulated surfaces, 
we introduce a length of the edges of the unit square (plaquette) $a$ and consider the continuum limit $a\to 0$ 
in approaching to the critical point. 

We put 
\be
\Delta=a^2t, \qquad  \lambda=\frac{3}{\sqrt{2}}\left(1+\frac12a^{3/2}\zeta\right),
\ee
where $t$ and $\zeta$ are interpreted as a (bulk) cosmological constant and a boundary cosmological constant in the continuum theory, respectively.   
Then, we obtain from (\ref{app:sol_omega}) 
\bea
\omega(\lambda) & = & \frac{11}{9\sqrt{2}}-\frac{8}{3\cdot 2^{1/4}}\,a \hat{\omega}(\zeta), \nn \\
\hat{\omega}(\zeta) & \equiv & t^{1/8}\sqrt{\zeta+t^{3/4}}-\frac{t^{1/2} }{2\cdot 2^{1/4}}+O(a^{1/2}).
\label{app:omega_hat}
\eea
$\hat{\omega}(\zeta)$ has a universal meaning in the critical behavior, which defines the quantity in the continuum theory.  
Here we find the unusual scaling of bulk and boundary cosmological constants implying $\vev{\mbox{(boundary length)}}\sim \vev{\mbox{(area)}}^{3/4}$ \cite{KZJ}.  

The one-point function $\vev{\frac{1}{N}\tr\left(M'XM'X\right)}$ is computed as 
\be
\vev{\frac{1}{N}\tr\left(M'XM'X\right)}_s' =  -\frac{2}{N^2}\frac{1}{Z_s'}\frac{\partial}{\partial g}Z_s' = -\frac{1}{N^2g}\vev{|m|}_s' 
=-\frac{1}{g}\left(\int^\alpha_0 dh\,h\rho(h)-\frac12\right)
\ee
at large $N$. 
By expanding (\ref{app:sol_H}) in large $h$, we find $\int^\alpha_0 dh\,h\rho(h)=-\frac16(2X^2+1)+X$. Finally, 
\be
\vev{\frac{1}{N}\tr\left(M'XM'X\right)}_s' =  \frac38-\frac{45}{16}a^2t+9a^3t^{3/2}+O(a^4).
\label{app:MXMX}
\ee
The third term is the leading non-analytic term at $t=0$, which is relevant to the critical behavior.   
The fractional power $t^{3/2}$ indicates that the string susceptibility exponent is $\gamma_{\rm str}=-1/2$ 
and the bulk surface is described by the same universality class as the $c=0$ pure gravity~\cite{KPZ,david,DK}. 

\section{Schwinger-Dyson equations}
\label{app:SDeqs}
\setcounter{equation}{0}
In this appendix, we derive several SD equations in the matrix model (\ref{app:S'_M_X}), by solving which we obtain one-point functions to be needed to compute a generalized EE including 
bulk gravity effects. 
Let $T^p$ ($p=1,\cdots, N^2$) a basis of $N\times N$ hermitian matrices satisfying 
\be
\tr\left(T^pT^q\right)=\delta^{pq},\qquad \sum_{p=1}^{N^2}\left(T^p\right)_{ij}\left(T^p\right)_{k\ell}=\delta_{i\ell}\delta_{jk}.
\ee
Matrices $M'$ and $X$ are expanded by the basis as
\be
M'=\sum_{p=1}^{N^2}M'_pT^p , \qquad X=\sum_{p=1}^{N^2}X_pT^p,
\ee
where $M'_p$ and $X_p$ are expansion coefficients. 

The SD equation 
\be
\vev{\frac{1}{N}\tr X^2}_s'=1-g\vev{\frac{1}{N}\tr\left(M'XM'X\right)}_s'  
\ee
is obtained from the identity 
\be
0=\int d^{N^2}M'\,d^{N^2}X\,\sum_{p=1}^{N^2}\frac{\partial}{\partial X^p}\,\left[\tr\left(T^pX\right)\,e^{-S_s'}\right].
\ee
Together with (\ref{app:MXMX}), we find 
\be
\vev{\frac{1}{N}\tr X^2}_s'=\frac{13}{12}-\frac34a^2t+2a^3t^{3/2}+O(a^4)
\label{app:XX}
\ee
in the large-$N$ limit.

By combining two SD equations from the identities
\bea
0& = & \int d^{N^2}M'\,d^{N^2}X\,\sum_{p=1}^{N^2}\frac{\partial}{\partial X^p}\,\left[\tr\left(T^p\frac{1}{z'-M'}X\right)\,e^{-S_s'}\right] , \nn \\
0 & = & \int d^{N^2}M'\,d^{N^2}X\,\sum_{p=1}^{N^2}\frac{\partial}{\partial M'^p}\,\left[\tr\left(T^p\frac{1}{z'-M'}\right)\,e^{-S_s'}\right] ,
\eea
we obtain 
\be
\omega_{X^2}(z')\equiv \vev{\frac{1}{N}\tr\left(\frac{1}{z'-M'}X^2\right)}_s'=\omega(z')-z'\left(\omega(z')^2-z'\omega(z')+1\right)
\label{app:omegaXX}
\ee
in the limit $N\to \infty$. 
Use of (\ref{app:omega_hat}) with 
$z'=\frac{3}{\sqrt{2}}\left(1+\frac12a^{3/2}\zeta'\right)$ 
yields 
\be
\omega_{X^2}(z')= \frac{40}{27\sqrt{2}}-\frac{44}{9\cdot 2^{1/4}}\,a\hat{\omega}(\zeta')+\frac{311}{108\sqrt{2}}\,a^{3/2}\zeta' -\frac{32}{34}\,a^2\hat{\omega}(\zeta')^2+O(a^{5/2}).
\label{app:omegaXX2}
\ee

The identity 
\be
0 = \int d^{N^2}M'\,d^{N^2}X\,\sum_{p=1}^{N^2}\frac{\partial}{\partial X^p}\,\left[\tr\left(T^p\frac{1}{z_1'-M'}X\frac{1}{z_2'-M'}\right)\,e^{-S_s'}\right] 
\ee
leads to the SD equation 
\begin{align}
\omega^{(2)}(z'_1,z'_2) \equiv & \vev{\left(\frac{1}{z'_1-M'}X\frac{1}{z_2'-M'}X\right)}_s' \nn \\
 = & \frac{1}{1+gz_1'z_2'}\left[\omega(z_1')\omega(z_2')+gz_1'\omega_{X^2}(z_1')+gz_2'\omega_{X^2}(z_2')-g\vev{\frac{1}{N}\tr X^2}_s'\right]
 \label{app:omega(2)}
\end{align}
at large $N$. Plugging into this (\ref{app:omega_hat}), (\ref{app:XX}), (\ref{app:omegaXX2}) and 
$z'_i= \frac{3}{\sqrt{2}}\left(1+\frac12a^{3/2}\zeta_i'\right)$ 
($i=1,2$), 
we see that both of the numerator and the denominator in (\ref{app:omega(2)}) start from the order of $a^{3/2}$ due to nontrivial cancellations. 
The result is  
\be
\omega^{(2)}(z_1',z_2') = \frac{391}{162}-\frac{64\sqrt{2}}{9}\,a^{1/2}\frac{\hat{\omega}(\zeta_1')\hat{\omega}(\zeta_2') +\hat{\omega}(\zeta_1')^2+\hat{\omega}(\zeta_2')^2-\frac{3}{4\sqrt{2}}t}{\zeta'_1+\zeta'_2}
+O(a).
\label{app:omega(2)2}
\ee
The second term is relevant to the critical behavior and provides the quantity in the continuum theory.

The definition of $\omega^{(2)}(z_1',z_2')$ is expanded in large $z_1'$ and $z_2'$ as 
\bea
\omega^{(2)}(z_1',z_2') & = & \sum_{k,\ell=0}^\infty\frac{1}{z_1'^{2k+1}z_2'^{2\ell+1}}\vev{\frac{1}{N}\tr\left(M'^{2k}XM'^{2\ell}X\right)}_s' \nn \\
& & +  \sum_{k,\ell=0}^\infty\frac{1}{z_1'^{2k+2}z_2'^{2\ell+2}}\vev{\frac{1}{N}\tr\left(M'^{2k+1}XM'^{2\ell+1}X\right)}_s' 
\eea
because of the $Z_2$ symmetry under $M'\to -M'$. 
The first term is extracted by 
\be
\omega^{(2)}_e(z_1',z_2') \equiv \frac12\left(\omega^{(2)}(z_1',z_2')-\omega^{(2)}(-z_1',z_2')\right)=z'_1z'_2\vev{\frac{1}{N}\tr\left(\frac{1}{z_1'^2-M'^2}X\frac{1}{z_2'^2-M'^2}X\right)}_s' .
\label{app:omega(2)e}
\ee
In the continuum limit, this becomes 
\be
\omega_e^{(2)}(z_1',z_2')=\frac56-\frac{32\sqrt{2}}{9}\,a^{1/2} \frac{\hat{\omega}(\zeta_1')\hat{\omega}(\zeta_2') +\hat{\omega}(\zeta_1')^2+\hat{\omega}(\zeta_2')^2-\frac{3}{4\sqrt{2}}t}{\zeta'_1+\zeta'_2}
+O(a),
\label{app:omega(2)e2}
\ee
whose second term is a half of that of $\omega^{(2)}(z',w')$ in (\ref{app:omega(2)2}).


\end{document}